\documentclass{aa}  

\usepackage{mathptmx}
\usepackage{txfonts}
\usepackage[T1]{fontenc}
\usepackage{ae,aecompl}

\usepackage{graphicx}	
\usepackage{mathtools,mathrsfs}
\usepackage{natbib}
\usepackage{hyperref}
\usepackage{ulem}
\usepackage{caption}
\usepackage{subcaption}
\usepackage{verbatim}
\usepackage{comment}

\newcommand{\gaeafid}{\textsc{GAEA}\xspace}
\newcommand{\tng}{\textsc{TNG}\xspace }
\newcommand{\btt}{\mathrm{B/T}}
\newcommand{\ob}{\textsc{OrbitBulge}\xspace}
\newcommand{\obdi}{\textsc{OBDI}\xspace}
\usepackage{xcolor}

\begin{document}

   \title{Incorporating Merger Orbit Effects into Bulge Formation in Semi-Analytic Model GAEA}
   \titlerunning{Reshape bulge formation in GAEA}
    \author{Lizhi Xie \inst{1} \corrauth{xielizhi.1988@gmail.com}
          \and 
          Lan Wang \inst{2,3} \email{wanglan@bao.ac.cn}
          \and
          Gabriella De Lucia\inst{4,5} \email{gabriella.delucia@inaf.it}
          \and
          Fabio Fontanot \inst{4,5}  \email{fabio.fontanot@inaf.it}
          \and 
          Guangquan Zeng \inst{6} \email{guangquan.zeng@outlook.com}
          }

    \institute{Tianjin Normal University, Binshuixidao 393, Tianjin, China 
    \and  National Astronomical Observatories, Chinese Academy of Sciences, Beijing 100101, China
    \and  School of Astronomy and Space Science, University of Chinese Academy of Sciences, Beijing 100049, China 
    \and  INAF – Astronomical Observatory of Trieste, via G.B. Tiepolo 11, I-34143 Trieste, Italy
   \and   IFPU - Institute for Fundamental Physics of the Universe, via Beirut 2, 34151, Trieste, Italy
   \and  Department of Physics, The Chinese University of Hong Kong, Sha Tin, N.T., Hong Kong, China     
   }

   \date{Received xxx; accepted xxx}
 



 \abstract{
 The morphological distribution of different galaxy population is tightly linked to the mix of their distinct merger histories, with the merger orbit playing an important role. }
 {In the framework of the GAEA semi-analytic galaxy formation model, we assess to what extent accounting for merger details can improve the model predicted galaxy morphologies.}
 {We develop a new algorithm (\ob) based on results of numerical simulations, that incorporates the dependence on collisional angle and gas fractions for bulge formation during mergers. We also adopt a new treatment of disc instability that accounts for instability in both gaseous and stellar discs and a reduced instability threshold. }
 {The new prescriptions promote bulge formation via minor mergers and suppress the contribution from major mergers consistently, therefore substantially increase the fraction of intermediate bulges across cosmic time. The revised disc-instability treatment further enhances bulge growth.}
 {These improvements lead to a better agreement with observed $\btt$ distribution with respect to the fiducial run. The new models predict that passive massive galaxies at $0<z<3$ are predominantly bulge-dominated, which is broadly consistent with observations. The predicted size-mass relations for star-forming and passive galaxies are also consistent with observational measurements at $z<1$. However, the updated models under-predict the bulge-dominated fraction for lower mass galaxies at high redshift, and lack of compact bulge-dominated galaxies, indicating the need for additional physics. }


\maketitle
\nolinenumbers



\section{Introduction}
\label{sec:introduction}

Morphology is one of the fundamental properties of galaxies. Providing a direct, instantaneous view of underlying orbital structure and dynamical state of galaxies \citep{zhu2018}, morphology is imprinted by the hierarchical formation history of the galaxy \citep{delucia2012}, and tightly correlates with other physical properties, such as stellar mass, star formation activity, and colors \citep{brinchmann2004,wuyts2011}. Understanding the evolution of galaxy morphology is one of the key questions in galaxy formation theory. 

Stellar components in a galaxy are broadly classified into the disc and the bulge (or spheroid). The disc is rotationally supported, while the bulge is dispersion-dominated. In the well known Hubble sequence \citep{hubble1926}, galaxies are divided by their visual morphology into early-type galaxies, with prominent bulges, and late-type galaxies, with obvious flat disc-like structure. The bulge-to-total ($\btt$)  ratio describes the prominence of the bulge in terms of mass/luminosity fraction enclosed in the bulge component, and represents a quantity that can be measured relatively easily for large photometric samples. It is widely used as a galaxy morphology indicator \citep{Simard2011}.
The Hubble sequence emerges as early as $z >3$ \citep{huertas-company2024, Genin2025}. Galaxies  at earlier times are found to be clumpy and irregular in structure. At cosmic noon, nearly $60 \%$ of galaxies exhibit disc features \citep{kartaltepe2023}. As time evolves, we observe an increasing fraction of bulge-dominated galaxies, which are largely quiescent \citep{lotz2008,huertas-company2025}. 

Many physical processes are involved in the bulge formation and evolution. Galaxy mergers, especially major mergers, play an important role in morphology transformation \citep{toomre1972, hopkins2010, delucia2011}. Bulge-dominated galaxies are preferentially found in dense environments, where merger rates were higher \citep{dressler1980}. 
The morphology of merger remnant galaxy depends on the properties of the merging system, including the mass ratio between the merging pair, gas fraction of the merging system, and orbital parameters, as pointed out by many simulations \citep{mihos1996, hopkins2009, moster2011}.
Secular evolution also plays an important role in bulge growth \citep{delucia2010, rodriguez-gomez2017}. Gas-rich galaxies at high redshift may suffer violent disk instabilities and form clumps that eventually sink to the galaxy center \citep{elmegreen2008}. In the framework of hydro-simulations, internal feedback and tidal heating are also found to exert non-negligible impact on the galaxy morphologies \citep{Scannapieco2008, zhang2025}, especially for low-mass galaxies. Their impact, however, depends on the sub-grid physical models and numerical resolution \citep{ludlow2023, yang2024, zeng2024}.

Semi-analytic models (SAMs) provide one of the major theoretical tools to study galaxy evolution, and are widely used for generating mock catalogs in preparation for large-scale surveys \citep{pei2024, tan2026}. Due to the lack of spatial resolution, SAMs adopt simplified algorithms designed  to account for the dominant physical processes driving bulge formation and morphological changes \citep{delucia2010}. Major mergers are typically assumed to form elliptical galaxies, whereas minor mergers do not destroy the disk and result in a slight increase in bulge mass per event \citep{kauffmann1993,baugh1996}. 
Stellar disc instability plays a minor role except for galaxies in the mass range of $\log M_{\star} \in $~[10.2,10.8] \citep{zoldan2018}. 
These simple prescriptions typically included in SAMs do not seem to capture well the morphological evolution following merger events or disc instabilities triggered by misaligned gas accretion from multiple direction \citep{wang2019, scannapieco2009}. Consequently, SAMs typically produce a strong bimodal $\btt$ distribution that peak around $\btt<0.2$ and $\btt> 0.9$ \citep{wilman2013, bluck2019}. In contrast, intermediate $\btt$ ratios are common in observation \citep{moffett2016}.

Several works have tried to address the tension between models and observations by incorporating more physical details responsible for morphological transformation. 
\citet{somerville2008} assumed that $20\%$ of disc stars are heated into the diffuse stellar halo. 
\citet{fontanot2015} derived a set of fitting formula from high-resolution hydro-dynamic simulation \citep{kannan2015}, in which the mass transferred to the remnant bulge depends on the mass ratio and gas fraction of the merger pairs. This approach successfully reduces the overproduction of massive elliptical galaxies. 
\citet{Irodotou2019} improved the agreement with observed fractions of morphological types via an improved treatment of disk instabilities. 
However, none of these studies considered the impact of merger orbits.

In recent years, cosmological hydrodynamical simulations with high resolution and large simulation volumes have been shown to exhibit a good agreement with observed galaxy morphology statistics \citep{tacchella2019,pfeffer2023}. These
enable statistical analysis of bulge formation via mergers. \citet{martin2018} found that the remnant morphology depends strongly on the inclination of the orbital plane against the central's and satellite's stellar disc, as well as on the gas fraction. \citet[][Z21]{zeng2021} analyzed major merger pairs in the \tng-100. They found that the remnant morphology tightly correlates with the collision angle $<\theta_{{\bf r},{\bf v}}>$, which is defined as the angle between the positional vector and the velocity vector averaged over $\sim1$ Gyr prior to merger. Satellite galaxies on a radial orbit, ($<\theta_{r,v}>\sim 0^\circ$), tend to lead to spheroid galaxies after major merger, while spiral-in mergers ($<\theta_{{\bf r},{\bf v}}>\sim 90^\circ$) tend to form disc galaxies. 
\citet{Wu2025} later found that the collision angle is a proxy for the alignment between the central and satellite galaxy relative to the orbital plane. 
Recent work \citep[][Z26]{zeng2026} extended the analysis to minor mergers and to lower-mass galaxies, and found the correlation between orbital angle and morphology transformation to still hold. A similar dependence is also found for galaxy pairs from EAGLE simulation (Gan et al. in preparation). 
These studies offer improved prescriptions that can be readily incorporated in a SAM framework. 

In this work, we incorporate the aforementioned correlation between merger orbits and remnant morphology into our semi-analytic model, together with a more physically motivated treatment for disc instabilities, motivated to improve the agreement with observed $\btt$ distribution. Our paper is structured as follows. Section~\ref{sec:model} introduces our fiducial model \gaeafid and the new modifications. In section~\ref{sec:results} we compare model predictions with observations, and discuss the impact of different bugle formation scenario. We then summarize our results and discuss the limitations of our work in section~\ref{sec:conclusion}.

\section{Modelling Bulge Formation}
\label{sec:model} 

In this work, we make use of the GAlaxy Evolution and Assembly semi-analytic model \citep[][GAEA]{delucia2024}. The model originates from the one presented in \citet{delucia2007}, but has been further developed to include a non-instantaneous stellar recycling \citep{delucia2014}, an updated supernovae feedback model based in part on sophisticated hydro-dynamical simulations \citep{hirschmann2016}, a self-consistent partition of cold gas into HI and H$_2$ \citep{xie2017}, environmental processes \citep{xie2020}, as well as an improved treatment for accretion onto super-massive black holes and AGN feedback \citep{fontanot2020}. Our reference model is calibrated to reproduce the evolution of stellar mass function in the redshift range $0 \le z < 3$ and AGN luminosity function in $0 \le z< 4$. The model predicts good agreement with the observed quenched fractions up to $ z\sim 3-4$, mass-metallicity relations out to $z\sim 3.5$, as well as other observational scaling relations from low to high redshift \citep{xie2018, fontanot2021, xie2024, fontanot2025, cantarella2025,osman2026}.

The model is run on the Millennium Simulation \citep[MS, ][]{Springel2005} and Millennium II Simulation \citep[MSII, ][]{boylan-kolchin2009} with a box size of $685$ and $137$ Mpc based on a WMAP 1-yr cosmology \citep{spergel2003} with $\Omega_m = 0.25$, $\Omega_b = 0.045$, $\sigma_8 = 0.9$, and $h = 0.73$. (add P-Millennium) The model adopts Chabrier IMF \citep{chabrier2003}.

In our semi-analytic model, each galaxy is composed of two stellar components, a disc and a bulge. We do not identify separately more detailed structure such as spiral arms, stellar haloes, and bars. 
The disc is assumed to follow an exponential density profile, and the bulge is assumed to follow Jaffe law \citep{jaffe1983}.

\subsection{Fiducial model}
\label{subsec:fid-model}

In \gaeafid, bulge grow via mergers and stellar disk instabilities. 

\subsubsection{Mergers}
\label{subsubsec: fid_merger}

Merger events are separated into `major' and `minor' mergers by the baryonic mass ratio between two progenitors, 
$\mu = \frac{M_{\star,sat}+M_{\rm cg,sat}}{M_{\star,{\rm cen}}+M_{\rm cg,cen}}=0.3$. Here $M_{\star}$ and $M_{\rm cg}$ are the stellar mass and cold gas mass. Quantities marked with ${\rm cen}$ and ${\rm sat}$ are for primary and secondary progenitor galaxies in a merger. 
Both major and minor merger can trigger a starburst, which transforms a fraction of cold gas into stars. The star burst ratio, which is the fraction of cold gas converted into stars during the burst, depends on the mass ratio \citep{somerville2001}
\begin{equation}
    e_{\rm burst} = 0.56 \mu^{0.7}.
    \label{eqn:sb}
\end{equation}
Then the total mass of newly formed stars is given by $M_{\rm burst} = e_{\rm burst}(M_{\rm cg, cen} + M_{\rm cg, sat})$.

During a major merger, all stellar components of the progenitor galaxies together with stars formed during the starburst form a new bulge. 
In a minor merger, the newly formed bulge comprises only the stars from the central galaxy's bulge and all stars from the satellite, while the disc of the central galaxy and the newly formed stars end up in the disc of the remnant galaxy. 
We can express the mass of the new bulge in a unified form as $M_{\rm b,new}=M_{\rm cen}+M_{\rm sat}$, where $M_{\rm cen}$ and $M_{\rm sat}$ are the total masses contributed by the central and satellite progenitors, respectively. These are computed as
\begin{equation}
M_{i} = f_{{\rm b},i}M_{{\rm b},i}
              + f_{{\rm d},i}M_{{\rm d},i}
              + f_{{\rm burst}} e_{\rm burst}M_{{\rm cg},i}.
\label{eqn:Mb}
\end{equation}
Here $i\in\{{\rm cen},{\rm sat}\}$ labels the two progenitor galaxies. 
$M_{{\rm b},i}$, $M_{{\rm d},i}$, and $M_{{\rm cg},i}$ denote the bulge, disc, and cold-gas masses of progenitor $i$, respectively.  
We define $f_{{\rm b},i}$ and $f_{{\rm d},i}$ as the fractions of bulge, disc components of progenitor $i$ that are deposited into the new bulge. The fraction $f_{{\rm burst}}$ represents the fraction of newly formed stars that is incorporated in the new bulge. 
In our fiducial implementation, all fractions are set to unity for major mergers, whereas $f_{\rm b, cen} =f_{\rm b, sat} = f_{\rm d, sat}=1$ and $f_{\rm d, cen}=f_{\rm burst}=0$ for minor mergers.

The corresponding mass-weighted radius of progenitor galaxy $i$ is estimated as
\begin{equation}
R_{i} = \frac{1}{M_i}\Big[
 f_{{\rm b}, i}M_{{\rm b}, i}R_{{\rm b}, i}
 + f_{{\rm d},i}M_{{\rm d},i}R_{{\rm d},i}
 + f_{{\rm sb},i} e_{\rm burst}M_{{\rm cg},i}R_{{\rm cg},i}\Big]
\label{eqn:Rb2}
\end{equation}
$R_{{\rm b},i}$, $R_{{\rm d},i}$ are the half-mass radii of bulge and disc. $R_{{\rm cg},i}$ is the half-mass radius of the cold gas disc, which we use as a proxy for the radius of the newly formed stars.

The bulge size after merger is computed assuming energy conservation, 
\begin{equation}
    C\frac{G M_{\rm b,new}^2}{R_{\rm b,new}} = C (1+C_{rad} f_{\rm cg})  (\frac{G M_{\rm cen}^2}{R_{\rm cen}} + \frac{G M_{\rm sat}^2}{R_{\rm sat}})  + \alpha_{inter} \frac{G M_{\rm cen}  M_{\rm sat}}{R_{\rm cen} + R_{\rm sat}} .
    \label{eqn:Rb}
\end{equation}
We adopt $C=0.5$ and $\alpha_{inter} = 0.5$ following \citet{guo2011}. Here $M_{\rm b,new}$ and $R_{\rm b,new}$ are the mass and half-mass radius of the new bulge. 
The parameter $C_{rad} f_{\rm cg}$ represents the fractional energy that is dissipated during merger \citep{covington2008}. The dissipation is a function of the gas fraction of the merger system, $f_{\rm cg}=\frac{M_{\rm cg, cen} + M_{\rm cg, sat}}{M_{\rm cen}+M_{\rm sat}}$. The fiducial model does not include energy radiative losses, and adopt $C_{rad} = 0$.

\subsubsection{disk instabilities}
\label{subsubsec:fid_di}

Following the criterion described in \citet{efstathiou1982}, the critical mass for a disc to maintain stability is  
\begin{equation}
    M_{\rm crit} =  \xi_{sdi}  \frac{V_{\rm max}^2 R_{\rm d}}{G}. 
    \label{eqn:DI}
\end{equation}
Here the $V_{\rm max}$ is the maximum circular velocity of the host halo. The empirical factor $\xi_{sdi} = 3$ is adopted to increase the instability threshold as suggested by simulations \citep{sellwood2001}. When the disc mass $M_{\rm d}$ exceeds the critical mass, the excess mass $ M_{\rm unstable} = M_{\rm d} - M_{\rm crit}$ is transferred from the disc to the bulge. We assume that the unstable stars transfer negligible angular momentum during this process, therefore the specific angular momentum of the disk becomes larger by a factor of $\frac{M_{\rm d}}{M_{\rm crit}}$. Consequently, the scale length of the stellar disk becomes larger, and the overall density of the stellar disc decreases. 

The size of the bulge after disk instability is 
\begin{equation}
    C\frac{G M_{\rm b,new}^2}{R_{\rm b,new}} = C (\frac{G M_1^2}{R_1} + \frac{G M_2^2}{R_2})  + \alpha_{inter} \frac{G M_1  M_2}{R_1 + R_2}.
    \label{eqn:Rb_inst}
\end{equation}
Here we use $\alpha_{inter}/C = 4$ following \citep{guo2011}.  $M_1$ and $R_1$ are the mass and half-mass radius of the original bulge. $M_2$ and $R_2$ are the mass and half-mass size of the unstable stars. We assume that unstable atars are located in the inner part of the disk, therefore the half-mass radius $R_{\rm unstable}$ can be estimated by
\begin{equation}
    M_{\rm unstable} =  M_{\rm d} \left[ 1 - (\frac{R_{\rm unstable}}{R_{\rm d}} + 1) exp(-\frac{R_{\rm unstable}}{R_{\rm d}}) \right].
\end{equation}

\citet[][and references therein]{kormendy2016} argued that bulge formed via secular evolution, i.e. disk instability, are more rotational-supported with S\'ersic indexes $n_S < 2$, differing from a dispersion-dominated classical bulge, which typically has $n_S\sim 4$ . 
In our model, we do not distinguish different bulge types, although we could keep track of the stars coming into the bulge through these different channels \citep{delucia2010}.

\subsection{Modified treatments}

\subsubsection{Dependence on orbit - \normalfont{\ob}}
\label{subsubsec:OB}

Inspired by the key dependencies for post-merger $\btt_{\rm after}$ extracted from the analysis of TNG merger pairs \citep{zeng2021, zeng2026}, we perform our own fit to these merger pairs, expressing $\btt_{\rm after}$ as a simple function of the $\btt$ of the primary galaxy $\btt_{\rm before}$, the baryonic mass ratio $\mu$, the average collision angle $<\theta_{{\bf r},{\bf v}}>$, and the mass of the primary galaxy $M_{\star, {\rm cen}}$,
\begin{equation}
    \btt_{\rm after} =  (a \mu +b) \cos <\theta_{{\bf r},{\bf v}}> + c  \log (M_{\star, {\rm cen}}/M_{\odot}) \btt_{\rm before} + d.
\label{eqn:fittingF}
\end{equation}
Here $a$, $b$, $c$, $d$ are free parameters.
It is important to note a few differences with respect to the work based on \tng, due to the different underlying framework.

\begin{table}[]
    \centering
    \begin{tabular}{ccccc}
                 & a & b & c & d \\
                 \hline
         gas-rich& 0.32 & 0.4 & 0.03 & 0.02\\
         \hline
         gas-poor& 0.24 & 0.32 & 0.05 & 0.12 \\ 
    \end{tabular}
    \caption{The parameters in equation~\ref{eqn:fittingF}.}
    \label{tab:fitParam}
\end{table}

First, we estimate the collision angle differently from Z21 and Z26. In GAEA, galaxies inherit positions and velocities directly from their host (sub)halos. This implies that we cannot trace accurately the orbital evolution of orphan galaxies, i.e. satellite galaxies whose host subhalos have been stripped below the resolution of the simulation. This implies that for most of our model galaxy pairs, we cannot measure the collision angle as in Z26 (i.e. averaging over 1Gyr before the merger). We thus  
estimate the average collision angle over the time interval when its parent subhalo can still be resolved, i.e. a few Gyrs prior to merger. If the host subhalo is destroyed immediately after infall, we continue tracking the orbit of the orphan satellite galaxy for at least 3 snapshots (that corresponds to $1 \pm 0.2$~Gyr at $z<2$). The orbit angle $\theta_{{\bf r},{\bf v}}$ continuously changes from infall time to prior-merger \citep{vasiliev2022, Pu2026}, and the presence of baryons in hydro-dynamic simulations also affects the merger orbits in a complicated manner \citep{Xu&Jing2025}. The difference between our estimate and that in Z26 is non-negligible, yet difficult to quantify. 

In addition, satellite galaxies keep evolving while approaching central galaxies, i.e. forming stars, consuming and recycling gas during star formation, and depleting gas due to ram pressure stripping. 
Satellite evolution also depends on the sub-grid physics adopted. Specifically, satellites in \tng deplete gas and quench more efficiently than in GAEA \citep{chen2024, xie2025}. Our model has neglected the dissipation and distortion the satellites stellar components, which are naturally included in \tng. 
Since our model does not resolve the  internal structure of merging galaxies, we are able to measure the $\btt$ ratios and mass ratios immediately pre/post-merger. Whereas for TNG, Z21 and Z26 measured the galaxy properties at $\sim 1$~Gyr pre/post-merger, when the interaction has not yet affected the internal structure of the merging galaxies and when the remnant's structure becomes stable. 
Given these discrepancies between the two approaches, we do not treat the merger pairs from \tng as a benchmark for equation~\ref{eqn:fittingF}. Rather, we independently recalibrate the parameters in equation~\ref{eqn:fittingF} to improve the agreement with the observational $\btt$ distribution for massive galaxies.

Since the gas fraction also influences the morphology of the merger remnant \citep{martin2018}, we treat gas-rich and gas-poor mergers separately, adopting a gas fraction threshold of $f_{\rm g}=\frac{M_{{\rm cg},cen}+M_{{\rm cg},cen}}{M_{\star,cen}+M_{\star,sat}} \sim 1$. We have verified that the final $\btt$ distribution is insensitive to this choice (e.g., varying the threshold to 0.2 yields negligible differences). The parameters included in the new formulation, and their reference values are summarized in Table~\ref{tab:fitParam}.  To minimize the effect of resolution, we apply the updated prescriptions to mergers with mass ratio $\mu > 0.001$. Satellite galaxies with lower mass ratios are directly added to the stellar disc of their centrals.

Using equation~\ref{eqn:fittingF} we compute the $\btt$ ratio of the remnant galaxy, and add stars from progenitor galaxies as well as the newly formed stars into the new bulge accordingly. All stars from the central progenitor's bulge are deposited into the new bulge. Should the bulge mass of the central progenitor galaxy be larger than the predicted mass, we ignore the negative growth and and assign only the central's bulge to the new bulge. On the other hand, if equation~\ref{eqn:fittingF} predicts a remnant bulge that is more massive than the central's bulge, we fill the new bulge with stars from satellite's bulge, satellite's disc, central's disc, and starburst stars in sequence, until the predicted mass is reached. All the remaining stars are assigned to the new disc of the remnant galaxy. We compute the contribution fractions from each reservoir ($f_{\rm b, sat}$, $f_{\rm d, sat}$, $f_{\rm d, cen}$, $f_{\rm burst}$) at this step, and use them to compute bulge size in equation~\ref{eqn:Rb2}, and to update the angular momentum of the new disc as introduced below.

Unlike the fiducial model, in which central stellar disc is either completely destroyed or undisturbed and maintains its angular momentum, the modified framework allows stars from satellite galaxies and starbursts to be partly accreted to the new disc, altering its angular momentum.
Assuming the angular momentum is conserved during merger, the angular momentum of the new stellar disk is, 
\begin{equation}
    \overrightarrow{L}_{\rm d,new} = (1-f_{\rm d, cen}) M_{\rm d,cen} \overrightarrow{j}_{\star, {\rm cen}} 
    +\overrightarrow{L}_{\rm ext}
    + (1-f_{\rm burst}) \overrightarrow{L}_{\rm burst}  .
    \label{eqn:L1}
\end{equation}
$M_{\rm d,cen}$ and $\overrightarrow{j}_{\star, cen}$ are the mass and specific angular momentum of the stellar disk of the central galaxy. 
The external angular momentum $\overrightarrow{L}_{\rm ext} $ carried by stars from the satellite galaxy is 
\begin{equation}
    \begin{split}    
    \overrightarrow{L}_{\rm ext} = & \big ((1-f_{\rm b, sat})  M_{\rm b, sat} +  (1-f_{\rm d, sat})       M_{\rm d, sat} \big ) \xi \overrightarrow{j}_{\rm orbit} \\ 
    &+ (1- f_{\rm d, sat})M_{\rm d, sat} \overrightarrow{j}_{\star, {\rm sat}},
    \end{split}    
    \label{eqn:L2}
\end{equation}
where $\overrightarrow{j}_{\rm orbit}$ and $\overrightarrow{j}_{\star, {\rm sat}}$ are specific angular momentum of the orbit and of the satellite's stellar disk. 
As discussed earlier, the orbit of orphan satellite galaxies cannot be reliably traced prior to merger, therefore we compute the orientation and modulus of the orbital plane when the host subhalo can still be resolved. A free parameter $\xi$ is adopted to account for the subsequent dissipation of the orbital angular momentum. In the following, we use $\xi = 0.1$. 

The angular momentum carried by the newly formed stars during starburst is 
\begin{equation}
    \overrightarrow{L}_{\rm burst} = (1-f_{\rm burst}) e_{\rm burst} (M_{\rm cg,cen} + M_{\rm cg,sat}) \overrightarrow{j}_{\rm cg, new}
    \label{eqn:L3}
\end{equation}
Here $\overrightarrow{j}_{\rm cg, new}$ is the specific angular momentum of the remanent galaxy's gas disk, which is 
\begin{equation}
    \overrightarrow{j}_{\rm cg,new} = \frac{M_{\rm cg,cen} \overrightarrow{j}_{\rm cg, cen} + M_{\rm cg,sat} (\overrightarrow{j}_{orbit} + \overrightarrow{j}_{\rm cg,sat}) }{M_{\rm cg, cen}  + M_{\rm cg,sat} }.
    \label{eqn:L4}
\end{equation}
We also consider the energy dissipation during gas-rich mergers by assuming $C_{rad} = 1$ in equation~\ref{eqn:Rb}. 
We implement the new merger-driven bulge formation algorithm and the updates in angular momentum in \gaeafid, and refer to this modified rendition as `\ob' (or `OB' in short) in the following. The new algorithm has a limited impact on the global galaxy properties, we slightly change a few parameters to recover the same level of agreement with the stellar mass functions and quenched fractions as in the reference run. The parameters and figures are shown in Appendix~\ref{appsec:calib}.

\subsubsection{Disc Instability - \normalfont{\obdi}}
\label{subsubsec:DI}

Low-mass galaxies ($M < 10^{10} M_{\odot}$) have lower merger rates and are thus less sensitive to modifications in the modeling of merger-driven bulge growth. Motivated to improve the agreement between $\btt$ distribution of galaxies in this mass range and observational constraints, we further revise our treatment of disc instabilities by including contribution from both gaseous and stellar disc within the framework of the \ob run. This run is referred as `\obdi' in the following.

After each star formation episode, we firstly examine the balance between gravity and centrifugal forces in the gaseous disc, then repeat the analysis for the stellar disc. Following the prescription adopted in \citet{Irodotou2019}. The critical mass of a stable gas disk is,
\begin{equation}
    M_{\rm crit, cg} =\frac{ V_{\rm max}^2 R_{\rm cg}}{ c^2_{\rm cg} G }. 
\end{equation}
Here $R_{\rm cg}$ is the scale length of the gas disk. The parameter $c_{\rm cg} = 0.9$ provides a  stability criteria for isolated gas disc \citep{christodoulou1995}. 
Excess gas in the inner disc is converted into stars,  and the remaining gas is redistributed to follow a exponential density profile. In our model the gas disc is divided into 21 annuli. The gas in the central region possesses lower specific angular momentum than gas at larger radii, as introduced in detail in \citet{xie2020}. Newly formed stars transfer angular momentum from the gaseous disc to the stellar disc corresponding to $\Sigma_i m_i \frac{r_i}{2r_{\rm cg}} \overrightarrow{j}_{\rm cg}$. Here $m_i$ is the mass of gas depleted from annulus $i$. $r_i$ is the radius of annulus $i$. $r_{\rm cg}$ and $\overrightarrow{j}_{\rm cg}$ are the scale length and the average specific angular momentum of the gaseous disc. After gaseous disc instability, both the effective radius and the average specific angular momentum of the remaining gas disc increase.

Instability in the gaseous disc leads to an accumulation of stars at the centre of the stellar disc. We then evaluate the stability of the stellar disc, as described in section~\ref{subsubsec:fid_di}, and transfer excess stars into the bulge. To increase the B/T ratio for low-mass galaxies, we make stellar disk instability more easily triggered by lowering the instability threshold in the stellar disk, adopting $\xi_{sdi} = 1.2$ in equation~\ref{eqn:DI}

In \gaeafid, stellar disc instability triggers a gas inflow that replenishes the accretion disk surrounding the central massive black hole and thence triggers quasar winds\citep[for details, we refer the readers to][]{fontanot2020}. In the modified model, we assume both gaseous and stellar disc instability could cause gas flowing to the centre, and the fraction of inflow gas is proportional to the mass of gas converted into stars during gaseous instability, as well as to the mass of stars that migrate to the bulge during stellar instability. The modified treatment leads to an enhancement in central black hole evolution that suppresses  subsequent star formation in massive galaxies. We then retune the parameters to reproduce the observed stellar mass functions and black hole-bulge mass relation at $z\sim 0$. The re-calibration is discussed in the Appendix~\ref{appsec:calib}. 
 
We note that the disc instability prescription is still oversimplified. By neglecting galactic bars, we do not consider their impact on the galactic discs \citep{athanassoula2008}. We also do not include gas inflows that could be triggered by the existence of galactic bars, which may lead to more violent disc instabilities. The connection between disc instabilities and central black hole accretion can also be refined. More detailed analysis on these issues, however, is beyond the scope of this paper. 

\section{Results}
\label{sec:results}

\subsection{Merger rates and contribution fraction}

\begin{figure*}
    \centering
    \includegraphics[width=\textwidth]{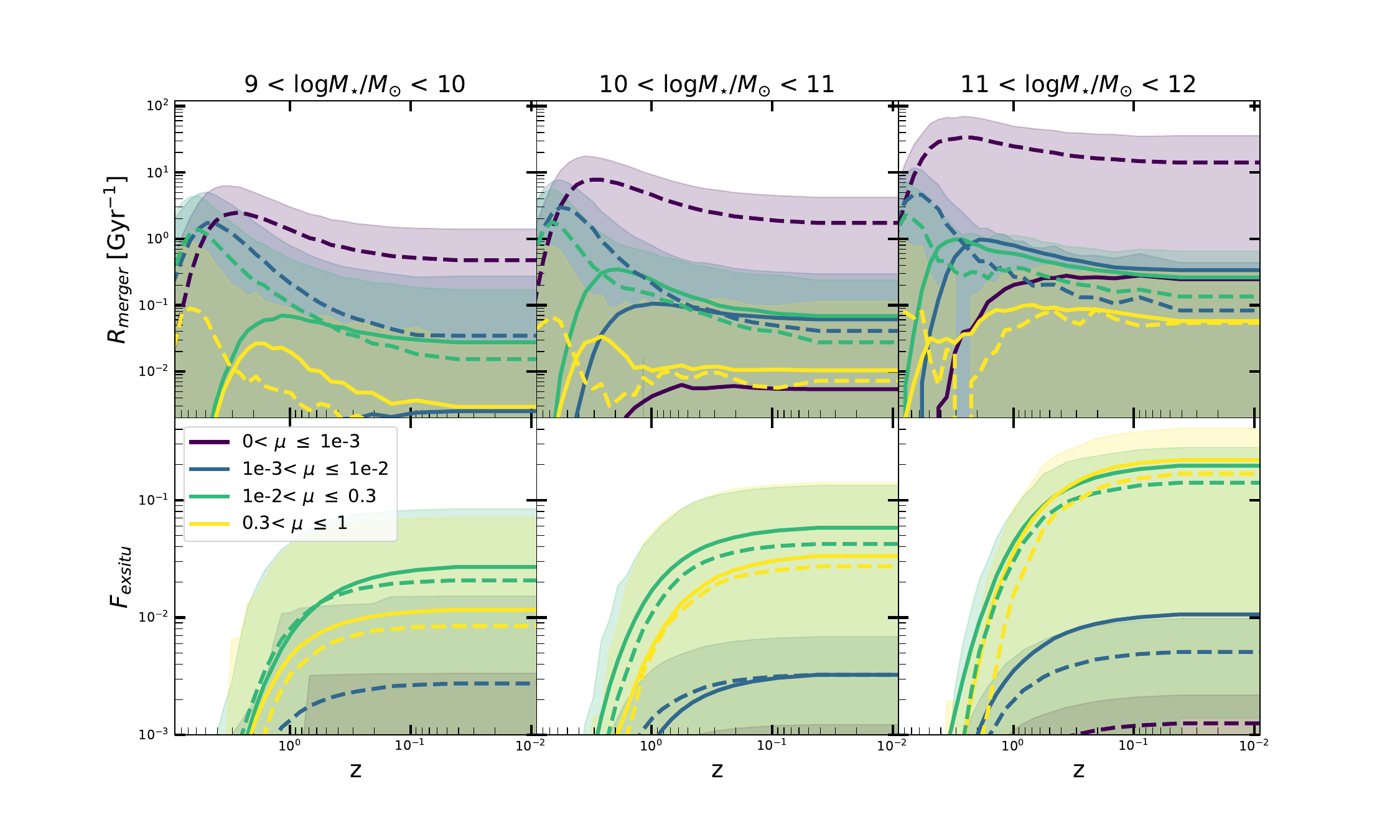}
    \caption{Top panels show the average merger rates of galaxies in different stellar mass bins as a function of redshift. Bottom panels show the cumulative contribution fraction of the merged galaxies to the final stellar mass of the primary galaxy. Different colors represent different mass ratios as labeled. Solid and dashed lines represent results for MS and MSII, respectively. The shaded regions are standard errors for the MSII galaxies.}
    \label{fig:mergerrates}
\end{figure*}

To investigate the role of different merger types in galaxy mass assembly, we present the merger rates and ex-situ mass fractions for galaxies at $z=0$ in three stellar mass bins.
The merger rate is defined as the number of merger events per unit time $R_{merger} = \frac{N_{merger}}{\delta t}$. The ex-situ mass fraction is defined as $F_{exsitu} = \frac{\Sigma_i M_{\star ,i}}{M_{\star, z=0}}$, where $\Sigma_i M_{\star ,i}$ is the total mass of merged satellite galaxies and $M_{\star, z=0}$ is the stellar mass at $z=0$. 
Results are shown in Figure~\ref{fig:mergerrates}. For simplicity, we classify mergers by their mass ratios: major ($\mu >0.3$), minor ($0.01 < \mu \leq 0.3$), mini  ($0.001 < \mu \leq 0.01$), and diffuse accretion ($\mu <0.001$). 

Overall, the merger rates and ex-situ mass fractions decrease with stellar mass. For galaxies more massive than $10^{11} M_{\odot}$, mergers contribute nearly half of the final stellar mass.  Both major and minor mergers contribute about 20 per cent each to the final stellar mass. For low-mass galaxies, however, most of the mass is formed in situ, with an ex-situ fraction less than 10 per cent. In these systems, minor mergers contribute more mass than major mergers. 
We also present results from the MSII simulation to examine resolution effects. MSII has a resolution 125 times higher than that of the MS. 
In all stellar mass bins we considered, the major and minor merger rates are consistent between MS and MSII at $z<1$, while at higher redshift the MSII resolves more mergers due to its higher resolution. The contribution fraction from major and minor mergers are roughly consistent too.

Mini mergers and diffuse accretion contribute much lower mass fractions than minor and major mergers, but they occur at a high rate. We find that excluding mini mergers from bulge growth in our modified \ob run produces an obvious reduction in $\btt$ ratios, whereas excluding diffuse accretion has a negligible impact on the final $\btt$ distribution. We therefore also include mini mergers in our treatment of merger-driven bulge formation. For massive galaxies ($M_{\star}>10^{10} M_{\odot}$), mini mergers are well resolved in the MS.  
For low-mass galaxies, both mini mergers and diffuse accretion are strongly suppressed in the MS compared to MSII. Low-mass galaxies in MSII experience one mini merger and some diffuse accretion on average, while the MS yields nearly zero for both. 
Therefore low-mass galaxies are affected by the limited resolution. We show the $\btt$ distribution of MSII in the appendix~\ref{appsec:MSII}. 

We note that GAEA predicts lower merger rates compared to TNG \citep{zakharova2026}. 
In GAEA the merger time is calculated immediately before the subhalo is destroyed, taking into account the mass ratios and position of subhalos \citep{Chandrasekhar1943}. Other important factors, e.g. dependence in orbital parameters, evolution in the halo and subhalo concentration, however, are neglected \citep{villalobos2013}. Furthermore, the presence of baryons complicates the issue and could either extend or shorten the merger time \citep{jiang2010, zhang2016}. These effects are not considered in our model. Comparing to the merger times fitting functions derived from TNG \citep{Xu&Jing2025}, merger times of galaxy pairs in GAEA are about 3 times longer. The discrepancy is even larger for mergers in low-mass halos. The lower merger rate in GAEA affects the morphology distribution of galaxies in our model. We plan to explore the topic in future work.

\subsection{Bulges at $z\sim 0$}
\label{subsec:btt_z0}

\begin{figure*}
    \centering
    \includegraphics[width=1.\textwidth]{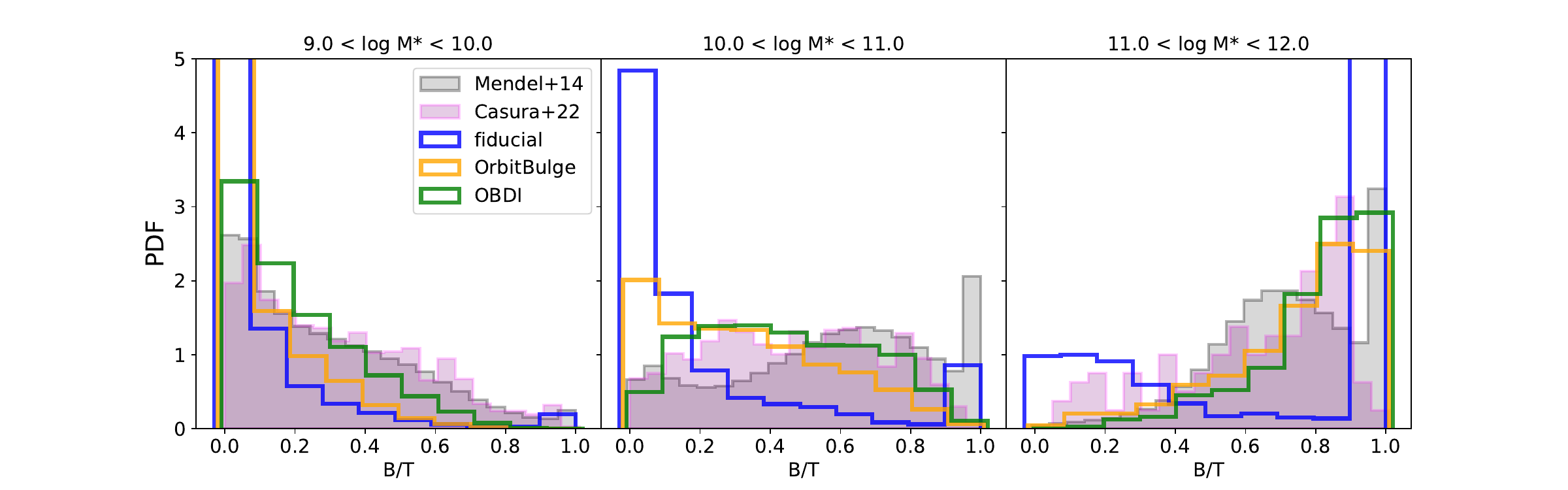}
    \caption{The $\btt$ distribution of galaxies at $z \sim 0$. Panels from left to right show the galaxies from low-mass to high-mass. Different colors correspond to predictions from different model renditions. Gray histograms show observational esimates based on SDSS galaxies from \citet{mendel2014} and GAMA+KiDS galaxies from \citet{casura2022}. }
    \label{fig:bttPDF_z0}
\end{figure*}

\begin{figure*}
    \centering
    \includegraphics[width=1.\textwidth]{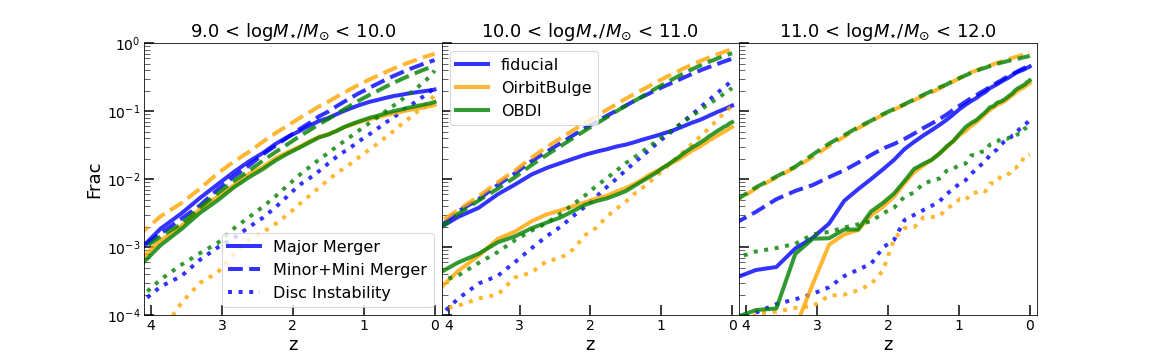}
    \caption{Fractional contribution to bulge mass assembly from different physical processes. Only galaxies that host a bulge are included. The solid, dashed, and dotted lines indicate the contributions from major mergers ($\mu > 0.33$), minor and mini mergers ($0.001< \mu \leq 0.33$), and disc instability, respectively. }
    \label{fig:bulgefrac}
\end{figure*}

We compare the predicted $\btt$  distributions for galaxies in the local Universe with observational measurements in Figure~\ref{fig:bttPDF_z0}. 
For observational results, we compare with the mass $\btt$ from the Sloan Digital Sky Survey (SDSS) Data Release 7 provided by \citet{mendel2014}. Based on the decomposition method \citep{Simard2011}, they analyzed $\sim657996$ galaxies and decompose them into a de Vaucouleurs bulge plus an exponential disk in five bands.  In the comparison, we select galaxies with reliable decomposition by excluding those labeled as type 4 in the original catalog. The stellar mass of each components - bulge, disc, and total stellar population - are derived via broadband SED fitting \citep{conroy2009} independently, assuming exponentially declining star formation history, Chabrier IMF \citep{chabrier2003}, and Calzetti (2000) extinction law \citep{calzetti2000}. Since in some cases the mass of bulge plus disc exceeds the total stellar mass, we consider the $\btt$ value as the ratio of bulge mass to the total mass of bulge and disc components for each galaxy in the sample. 

We also compare with the luminosity $\btt$ ratio from \citet{casura2022}, which provides structural parameters for $13096$ galaxies at $z < 0.08$ from the Galaxy And Mass Assembly survey \citep{driver2011} and Kilo-Degree Survey \citep[KiDS;][]{dejong2013}. The stellar masses are derived by Bayesian SED fitting of aperture-matched 21-band photometry \citet{taylor2011}, assuming exponentially declining star formation histories, the stellar population models of \citet{bruzual2003}, a \citet{chabrier2003} IMF, and the \citet{calzetti2000} extinction law. 
We select galaxies that are fitted with a S\'ersic bulge plus an exponential disc to the r-band surface brightness profile for reliable $\btt$ measurements, which leaves only 3210 galaxies. Our selection has excluded about 65, 20, 10 per cent of galaxies from the low-mass, intermediate mass, massive stellar mass bins, most of which prefer a single S\'ersic fit. Including these galaxies would significantly increase the fraction of low-mass disc, low-mass elliptical, and massive elliptical galaxies, as shown in Fig~\ref{fig:btt_gasdi}.  

Compared with the broad $\btt$ distribution of observational galaxies (gray and magenta histograms in Fig~\ref{fig:bttPDF_z0}), our fiducial model (blue histograms) predicts a strong bimodality across all stellar mass bins, with clear peaks corresponding to elliptical galaxies ($\btt > 0.9$) and disc-dominated galaxies ($\btt < 0.2$), and a lack of intermediate bulges that are common in observation.  

The implementation of the \ob algorithm (orange histograms), introduced in sec~\ref{subsubsec:OB}, eliminates the bimodal $\btt$ distribution in the fiducial model, and produces more intermediate bulges. The result is expected, as the $\btt$ of remnant galaxy depends on the collision angle, which exhibits a broad distribution from $0^\circ$ to $90^\circ$. 
In Figure~\ref{fig:bulgefrac} we plot the fractional contribution to the final bulge mass of each physical process. The new algorithm reduces the contribution to bulge growth from major mergers, and increases the contribution from minor and mini mergers. 
The $\btt$ distribution predicted by the new model are in better agreement with observational measurements for the more massive bins. At lower masses, since the rates of mini mergers are under-estimated in the MS simulation, their $\btt$ distribution is only slightly affected by new algorithm, and their bulge ratios are lower with respect to observational estimates. 

The \obdi run (green histograms) reproduces the observed $\btt$ better, particularly for low-mass galaxies. The improvement is mainly ascribed to the lower instability threshold for the stellar disk, while gaseous disc instability plays a minor role (as shown in Fig~\ref{fig:btt_gasdi}). In the \obdi run, the fractional contribution from disk instability to bulge growth increases across all stellar masses and redshifts. But its contribution remains lower than the minor and mini mergers (Figure~\ref{fig:bulgefrac}). In our modified model, minor and mini mergers are the dominant channel to grow bulges. 

In summary, considering the impact of merger orbits on remnant galaxies' morphology brings model predictions to a better agreement with observations for massive galaxies. Modifying the disc instability can further improve model predictions for low-mass galaxies. Since the merger orbits inherit the tidal torque of the large-scale environment, this \ob algorithm potentially affects the dependence of the morphology distribution on the cosmic-web. We have investigated this topic in a dedicated paper (Xu et al. in prep.)

\subsection{The $\btt$ distribution at higher redshifts}

\label{subsec:btt_highz}
\begin{figure*}
    \centering
    \includegraphics[width=1.\textwidth]{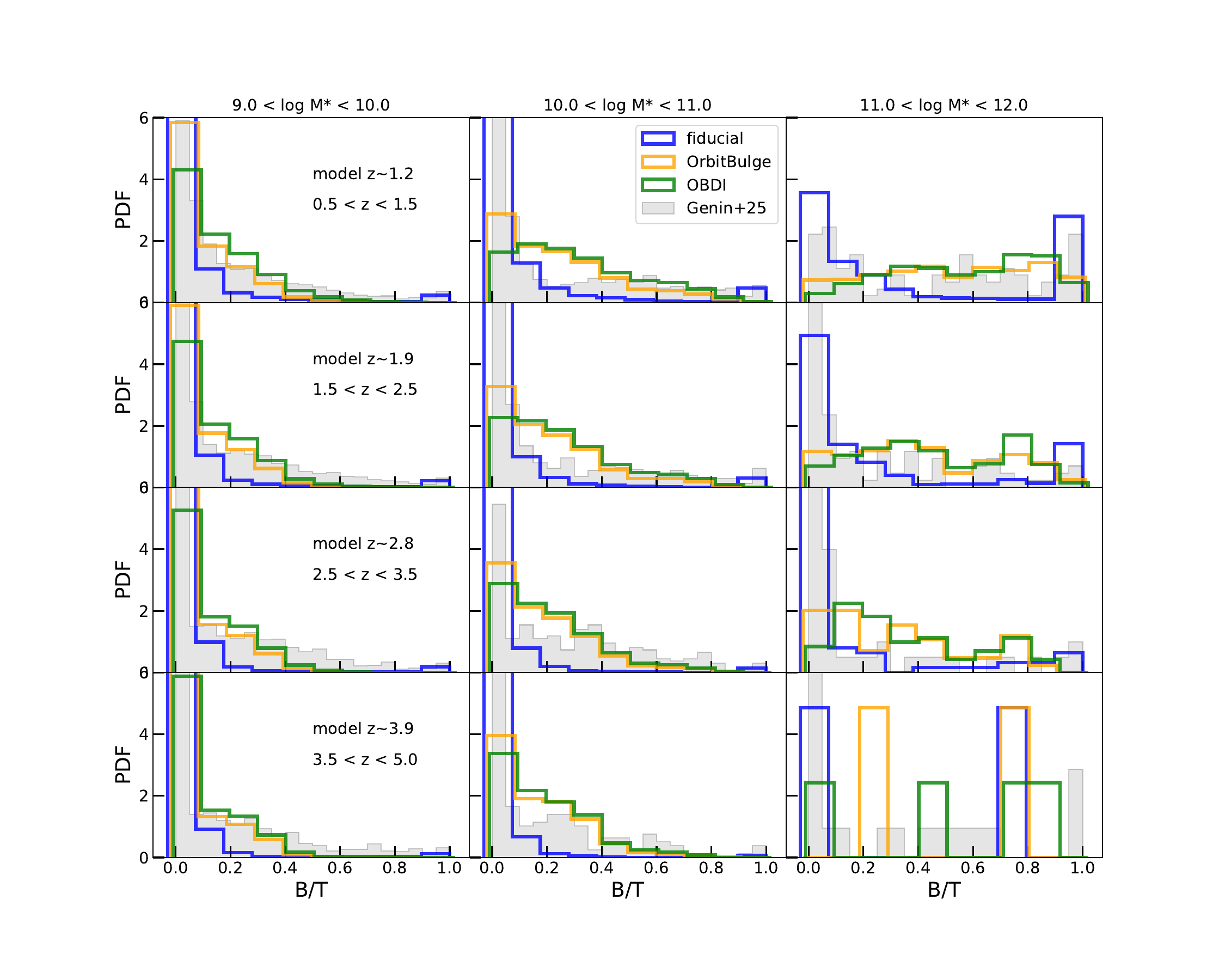}
    \caption{The $\btt$ distribution of galaxies at higer redshift. Panels from left to right show the galaxies from low-mass to high-mass. Different colors correspond to predictions from different model renditions. The observational measurements are from the DAWN JWST Archive \citep{Genin2025}. We use the F200W band at $z<2.5$, and F356W at higher redshift. }
    \label{fig:bttPDF_highz}
\end{figure*}

We then investigate how the different physical prescriptions affect bulge formation at earlier cosmic epochs. Figure~\ref{fig:bttPDF_highz} shows the $\btt$ distribution out to redshift $z\sim 5$. 
We compare with observational measurements from the DAWN JWST Archive (DJA) \footnote{https://dawn-cph.github.io/dja/blog/2024/08/16/morphological-data/} \citep{Genin2025}, which provides bulge-disk decomposition (B+D) for 67035 galaxies observed with NIRCam onboard JWST. Observational estimates have been obtained by fitting the multi-bands brightness profiles with an exponential disk profile and a de Vaucouleurs bulge profile with S\'ersic index $n_S =4$. We only use galaxies with flag$=2$ that correspond to robust fitting result. This leaves 59046 galaxies. There are 19191, 16311, 9599, 7290 galaxies in the redshift ranges of [0.5,1.5), [1.5,2.5), [2.5,3.5), [3.5,5). In this work, we use the B+D fitting in the $F200W$ band for galaxies at $z<2.5$, and $F356W$ band at higher redshifts. We note that the galaxy sample is not complete, and may have large uncertainties in the measurements of decomposition and photometric redshift. Our comparisons should therefore be considered with caution. 

Bulges emerge at very early time in the DJA sample. Bulge-dominated galaxies with $\btt > 0.5$ are discovered across all stellar mass bins at $3.5 \le z < 5$. Some of them are even elliptical galaxies with $\btt> 0.9$. Massive galaxies are more bulge-dominated than lower mass galaxies at all redshifts. For massive galaxies with $M_{\star} > 10^{10} M_\odot$, the fraction of bulge-dominated galaxies increases as time evolves.  For low-mass galaxies, there is no significant evolution as a function of cosmic time.  

In our fiducial model, there is a small fraction of elliptical galaxies emerging at $z> 3.5$, which are remnants of major mergers. The implementation of \ob reduces the fraction of elliptical galaxies at early epochs, however, enhancing bulge formation in overall galaxy population. There are more galaxies hosting bulges in the updated models, though the bulge mass ratios are lower than the observational values. The modified disc instability recipe slightly increases the bulge fractions since $z\sim 5$. The impact becomes more obvious at $z\sim 1$. In summary, the modified bulge formation recipe overall produces larger $\btt$ ratios than the fiducial model, which improves the agreement with observation. However, it fails to produce the large numbers of bulge dominated galaxies that are discovered in the early Universe, pointing out the necessity to include additional physical processes.

\subsection{The correlation between star formation activity and morphology}

\begin{figure*}
    \centering
    \includegraphics[width=1.\textwidth]{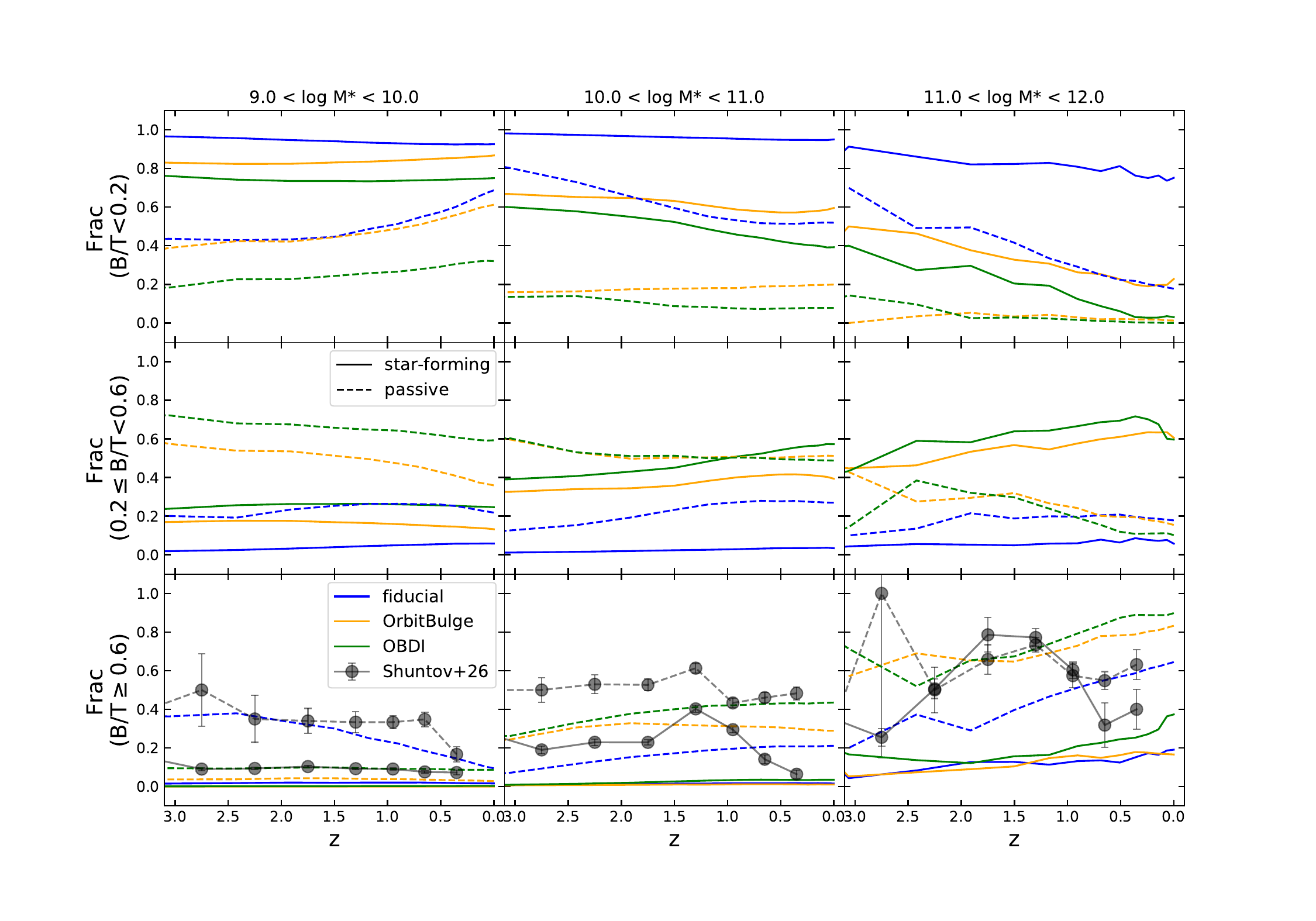}
    \caption{The redshift evolution of fractions of disc-dominated ($\btt<0.2$), intermediate bulge ($0.2\leq < \btt<0.6$), and bulge-dominated ($\btt \geq 0.6$) galaxies among star-forming and passive galaxies. Star-forming (solid lines) and passive (dashed lines) galaxies are divided by the threshold $0.3/t_{\rm H}$, where $t_{\rm H}$ is the Hubble time. The blue, orange, and green curves correspond to the fiducial, \ob, and \obdi runs, respectively.}
    \label{fig:FracBTT_z}
\end{figure*}

In this section, we further investigate whether bulges grow primarily in star-forming or quiescent galaxies. Figure \ref{fig:FracBTT_z} illustrates the fractions of galaxies in given $\btt$ bins among star-forming (solid lines) and quiescent (dashed lines) galaxies. 
Galaxy are classified into three types from top to bottom: disk-dominated ($\btt < 0.2$), intermediate ($0.2 \le \btt < 0.6$), and bulge-dominated ($\btt \ge 0.6$) systems. 

The figure reveals a clear dependence of galaxy morphology on star formation activity. For low-mass galaxies shown in the left column, star-forming galaxies are predominantly disc-dominated across all redshifts, while quiescent galaxies show a mixture of morphological types. In the fiducial run, about $40\%$ of passive low-mass galaxies are disc-dominated at $z\sim 3$, and another $\sim 40\%$ are bulge-dominated. Whereas in the new models, most passive galaxies host intermediate bulges. The disc-dominated fraction increases towards lower redshift. The evolutionary trend indicates that quenching in low-mass galaxies at high redshift is tightly correlated with bulge formation processes, i.e. merger and disc-instability; whereas it is dominated by environmental effects at lower redshift, which does not directly affect galaxy morphology in our model, though ram pressure stripping may enhance the probability of gas disc instabilities by stripping gas from the outer disc. 

The models diverge more significantly for more massive galaxies. In the fiducial model, most star-forming massive galaxies are disc-dominated across all redshifts. The fraction of bulge-dominated morphology slowly increases and reach $20 \%$ at $z=0$. In the updated models, a large fraction of star-forming galaxies host intermediate bulges, and the fraction increases as time evolves. At $z=0$, about $50\%$ and $80\%$  of star-forming galaxies with $\log M_* > 11 M_{\odot}$ possess intermediate bulges in the \ob and \obdi model. 
Morphology of passive massive galaxies are also strongly altered by the new bulge formation prescriptions. In the fiducial model, they are predominantly disc-like galaxies at $z\sim 3$. The bulge-dominated passive population emerges at later time. At $z=0$, about $20 \%$ of passive massive galaxies are still disc-dominated. We pointed out in previous work that massive galaxies at high redshifts are quenched by strong quasar winds, likely driven by minor mergers \citep{xie2024, delucia2024}. Consequently, they are quenched while maintaining their disc shape. In the new runs, minor mergers can create massive bulges, so high-redshift passive massive galaxies are mostly bulge-dominated, with a small fraction possessing intermediate bulges.
As bulges grow with time, more than $80\%$ of passive massive galaxies in the new runs attain $\btt > 0.6$ at low redshift.

The dependence of morphology on star formation activity predicted by our models is broadly consistent with observed trends, with passive galaxies hosting more massive bulges than star-forming galaxies \citep{wuyts2011, Genin2025}. More specifically, we compare with the bulge-dominated ($\btt > 0.6$) fractions of the quiescent and star-forming population from \citet{shuntov2026}. They analyzed over 700,000 galaxies from COSMOS2025, divided them into star-forming and passive using NUVrJ color, and fit the flux $\btt$ ratio with a $n_S =4 $ component and a $n_S =1$ component \citep{shuntov2025}.
Observationally, more than half of passive massive galaxies are bulge-dominated, and our new runs reproduce this result well. Similar results were reported in \citet{kawinwanichakij2026}, who studied 17 spectroscopically confirmed quiescent massive galaxies ($10.2<\log M_{\star}/M_{\odot} < 1.2$) at $3<z<4.3$, and found most of them are bulge-dominated with only 2 exhibiting disc-dominated morphology ($\btt \sim 0.2$).  
Although number statistics are low and uncertainties are potentially large, our new models are overall more consistent with the observed trends than the fiducial run. 

Model predictions differ from observed morphological fractions for star-forming galaxies and passive low-mass galaxies at high redshifts. In \citet{shuntov2026}, more than half star-forming massive galaxies at $1.5 < z < 2$ are bulge-dominated. \citet{huertas-company2025} analyzed the morphology of the COSMOS2025 sample using Convolutional Neural Network and found comparable number densities of disc-dominated and bulge-dominated morphologies among massive star-forming galaxies at this redshift.
Our new models predict larger $\btt$ values for massive star-forming galaxies than the fiducial run, yielding more massive bulges and thus better agreement with these observations. Nevertheless, these galaxies still host only intermediate bulges in our models. For the passive galaxies at high redshifts in the low and intermediate mass bins, about $40-50\%$ of them are bulge-dominated in \citet{shuntov2026}, and even higher bulge-dominated and peculiar fractions are suggested by machine learning classification \citet{liu2021, huertas-company2025}. Our new models under-predict the bulge-dominated fractions, , which may suggest that some relevant physics is missing. 
However, given the large uncertainties in the observational measurements, drawing robust conclusions might be premature.

\subsection{Size-mass relation}
\begin{figure*}
    \centering
    \includegraphics[width=1.\textwidth]{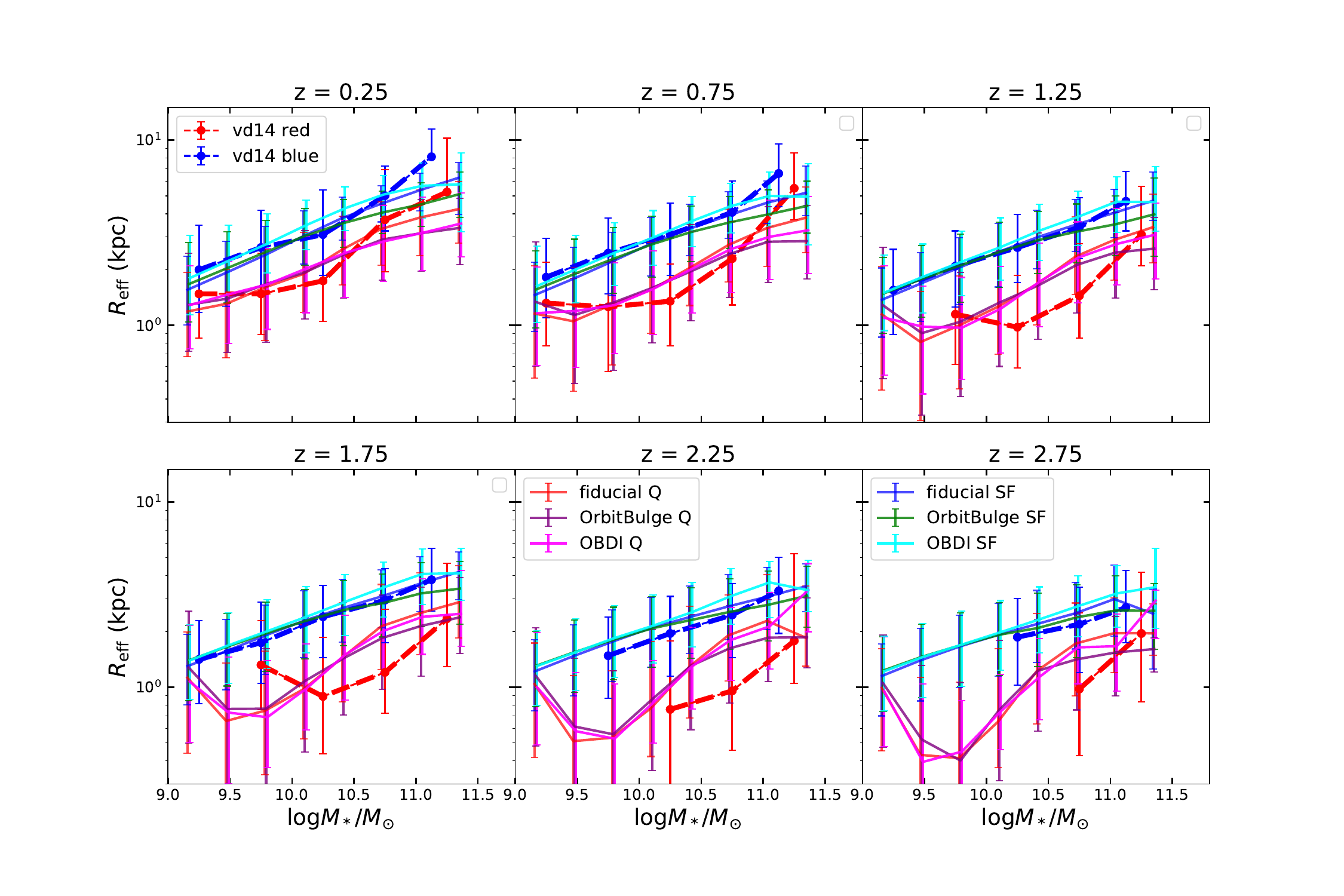}
    \caption{The redshift evolution of median size mass relation for star-forming and quiescent galaxies. Error bars are 16, 84 per cent scatters. The observational measurements are from \citet{vanderwel2014}.}
    \label{fig:Re-mass}
\end{figure*}

Galaxy sizes provide an additional and independent constraint. Combining 3D-HST survey and CANDELS imaging data,  \citet{vanderwel2014} found that galaxy sizes depend on their color and stellar mass. When dividing galaxies into late-type and early-type by UVJ colors, they pointed out that the size-mass relation of early-type galaxies grow rapidly since $z\sim 3$, whereas late-type galaxies evolve moderately. 
In Figure~\ref{fig:Re-mass} we compare the model-predicted size-mass relations with observations. We divide galaxies into star-forming and passive by $sSFR \sim 0.3/t_{\rm H}$, and compute the projected half-mass radius for each model galaxy. We compare model predictions against observational measurements for the circularized effective radii corrected to rest-frame $5000 \AA$ from \citet{vanderwel2014}. 

All models reproduce the observed size-mass relations for star-forming galaxies at $0<z<3$. The \obdi run predicts slightly larger effective radii than the fiducial model at fixed stellar mass, mainly because disc instability increases the specific angular momentum of stellar disc. The \ob run yields smaller radii for massive star-forming galaxies, as these galaxies have larger $\btt$ ratios than their counterparts in the fiducial model.  
For passive galaxies, the model-predicted size-mass relations are broadly consistent with observations at $z<1$, but are above the observed relations at $z> 1$, especially in the stellar mass range of $10< \log M_{\star}/M_{\odot} <11$.  
The \ob and \obdi versions incorporate radiative energy loss during gas-rich mergers as indicated in Section~\ref{subsubsec:OB}, which reduces the effective radius by about 0.1 dex for galaxies with $\log M_{\star}/M_{\odot} \sim 10.5$ at $z\sim 2$. Nonetheless, the new models still over-predict the sizes for passive galaxies.

\section{Summary}
\label{sec:conclusion}

In this work, we test how the morphological distribution of galaxies from semi-analytic model can be improved by incorporating more physical details responsible for bulge formation. More specifically, we 1) develop a new algorithm that accounts for bulge formation during mergers by considering the impact of merger orbits and gas fractions (\ob); 2) update the treatments of instabilities in stellar and gaseous discs (\obdi). 

The \ob algorithm hinders bulge formation during major mergers and boosts bulge formation during minor mergers in a consistent manner, which leads to a broad distribution of $\btt$ ratios for model galaxies. The improved DI prescription further increases the $\btt$ ratios especially for low-mass galaxies. We show that model realizations implementing such new prescriptions result in an improved agreement with the $\btt$ distribution and size-mass relation for star-forming and quiescent galaxies at low redshifts. 

Compared to the fiducial run, galaxies in new models generally have higher bulge ratios, especially massive galaxies at higher redshifts. In updated models, the fraction of massive galaxies with pure disc morphology is significantly reduced, while the fraction of intermediate bulges ($0.2 < \btt < 0.6$) increases. A large fraction ($\sim 50 \%$) of star-forming massive galaxies host intermediate bulges, while almost all passive massive galaxies host intermediate or more massive bulges. These predictions can be tested against upcoming JWST observations. 

Our new models, however, fail to reproduce the small sizes of passive galaxies at high redshift ($z>1$), even when energy dissipation in gas-rich mergers is included. This discrepancy is partly ascribed to the lack of elliptical galaxies in our model, and partly to over-prediction of bulge and disc sizes. The disagreement between model and observation indicate that the physical prescription for bulge formation at high redshift is still inadequate and requires further revision.
\citet[][and references therein]{Dekel&Burkert2014} pointed out that the gas inflow from cosmic web at high redshift may trigger violent disc instabilities, driving dissipative contraction that transforms star-forming galaxies into a phase called `blue nuggets'. Our model does not include this process, and this might be the reason for the reported lack of galaxies that are star-forming, spheroidal, and compact. Furthermore, internal feedback from supernovae and AGN could destroy galactic discs and induce turbulence that enables the formation of bulges \citep{lehnert2014, benavides2025}. We plan to test these hypotheses in future work.

\section*{Acknowledgements}
LZX acknowledges support from the Ministry of Science and Technology of China (grant No.2020SKA0110100), the National Natural Science Foundation of China (NSFC, grand number 12041302). 
LW acknowledges the supports from NSFC (grant No.~12588202) , the National SKA Program of China (No. 2022SKA0110201), and CAS Project for Young Scientists in Basic Research No. YSBR-062.

\section*{Data availability}
The data underlying this article will be shared on reasonable request to the corresponding author. 



\bibliographystyle{aa}
\bibliography{references} 

@ARTICLE{athanassoula2008,
       author = {{Athanassoula}, E.},
        title = "{Disc instabilities and semi-analytic modelling of galaxy formation}",
      journal = {\mnras},
         year = 2008,
        month = oct,
       volume = {390},
       number = {1},
        pages = {L69-L72},
          doi = {10.1111/j.1745-3933.2008.00541.x},
archivePrefix = {arXiv},
       eprint = {0808.0016},
 primaryClass = {astro-ph},
       adsurl = {https://ui.adsabs.harvard.edu/abs/2008MNRAS.390L..69A}
}

@ARTICLE{baugh1996,
       author = {{Baugh}, C.~M. and {Cole}, S. and {Frenk}, C.~S.},
        title = "{Evolution of the Hubble sequence in hierarchical models for galaxy formation}",
      journal = {\mnras},
         year = 1996,
        month = dec,
       volume = {283},
       number = {4},
        pages = {1361-1378},
          doi = {10.1093/mnras/283.4.1361},
archivePrefix = {arXiv},
       eprint = {astro-ph/9602085},
 primaryClass = {astro-ph},
       adsurl = {https://ui.adsabs.harvard.edu/abs/1996MNRAS.283.1361B}
}

@ARTICLE{boylan-kolchin2009,
       author = {{Boylan-Kolchin}, Michael and {Springel}, Volker and {White}, Simon D.~M. and {Jenkins}, Adrian and {Lemson}, Gerard},
        title = "{Resolving cosmic structure formation with the Millennium-II Simulation}",
      journal = {\mnras},
         year = 2009,
        month = sep,
       volume = {398},
       number = {3},
        pages = {1150-1164},
          doi = {10.1111/j.1365-2966.2009.15191.x},
archivePrefix = {arXiv},
       eprint = {0903.3041},
 primaryClass = {astro-ph.CO},
       adsurl = {https://ui.adsabs.harvard.edu/abs/2009MNRAS.398.1150B}
}

@ARTICLE{benavides2025,
       author = {{Benavides}, Jos{\'e} A. and {Sales}, Laura V. and {Wetzel}, Andrew and {Moreno}, Jorge and {Feldmann}, Robert and {Mercado}, Francisco J. and {Bullock}, James S. and {Hopkins}, Philip F. and {Faucher-Gigu{\'e}re}, Claude-Andr{\'e} and {Stern}, Jonathan and {Wheeler}, Coral and {Kere{\v{s}}}, Du{\v{s}}an},
        title = "{Disks no more: the morphology of low-mass simulated galaxies in FIREbox}",
      journal = {\mnras},
         year = 2025,
        month = oct,
       volume = {544},
       number = {4},
        pages = {4651-4664},
          doi = {10.1093/mnras/staf1847},
archivePrefix = {arXiv},
       eprint = {2508.00991},
 primaryClass = {astro-ph.GA},
       adsurl = {https://ui.adsabs.harvard.edu/abs/2025MNRAS.544.4651B}
}

@ARTICLE{brinchmann2004,
       author = {{Brinchmann}, J. and {Charlot}, S. and {White}, S.~D.~M. and {Tremonti}, C. and {Kauffmann}, G. and {Heckman}, T. and {Brinkmann}, J.},
        title = "{The physical properties of star-forming galaxies in the low-redshift Universe}",
      journal = {\mnras},
         year = 2004,
        month = jul,
       volume = {351},
       number = {4},
        pages = {1151-1179},
          doi = {10.1111/j.1365-2966.2004.07881.x},
archivePrefix = {arXiv},
       eprint = {astro-ph/0311060},
 primaryClass = {astro-ph},
       adsurl = {https://ui.adsabs.harvard.edu/abs/2004MNRAS.351.1151B}
}

@ARTICLE{bluck2019,
       author = {{Bluck}, Asa F.~L. and {Bottrell}, Connor and {Teimoorinia}, Hossen and {Henriques}, Bruno M.~B. and {Mendel}, J. Trevor and {Ellison}, Sara L. and {Thanjavur}, Karun and {Simard}, Luc and {Patton}, David R. and {Conselice}, Christopher J. and et al.},
        title = "{What shapes a galaxy? - unraveling the role of mass, environment, and star formation in forming galactic structure}",
      journal = {\mnras},
         year = 2019,
        month = may,
       volume = {485},
       number = {1},
        pages = {666-696},
          doi = {10.1093/mnras/stz363},
archivePrefix = {arXiv},
       eprint = {1902.01665},
 primaryClass = {astro-ph.GA},
       adsurl = {https://ui.adsabs.harvard.edu/abs/2019MNRAS.485..666B}
}

@ARTICLE{bruzual2003,
       author = {{Bruzual}, G. and {Charlot}, S.},
        title = "{Stellar population synthesis at the resolution of 2003}",
      journal = {\mnras},
         year = 2003,
        month = oct,
       volume = {344},
       number = {4},
        pages = {1000-1028},
          doi = {10.1046/j.1365-8711.2003.06897.x},
archivePrefix = {arXiv},
       eprint = {astro-ph/0309134},
 primaryClass = {astro-ph},
       adsurl = {https://ui.adsabs.harvard.edu/abs/2003MNRAS.344.1000B}
}

@ARTICLE{calzetti2000,
       author = {{Calzetti}, Daniela and {Armus}, Lee and {Bohlin}, Ralph C. and {Kinney}, Anne L. and {Koornneef}, Jan and {Storchi-Bergmann}, Thaisa},
        title = "{The Dust Content and Opacity of Actively Star-forming Galaxies}",
      journal = {\apj},
         year = 2000,
        month = apr,
       volume = {533},
       number = {2},
        pages = {682-695},
          doi = {10.1086/308692},
archivePrefix = {arXiv},
       eprint = {astro-ph/9911459},
 primaryClass = {astro-ph},
       adsurl = {https://ui.adsabs.harvard.edu/abs/2000ApJ...533..682C}
}

@ARTICLE{cantarella2025,
       author = {{Cantarella}, Sebastiano and {De Lucia}, Gabriella and {Fontanot}, Fabio and {Hirschmann}, Michaela and {Xie}, Lizhi and {Franco}, Maximilien and {Plat}, Ad{\`e}le},
        title = "{Probing the dawn of galaxies: star formation and feedback in the JWST era through the GAEA model}",
      journal = {arXiv e-prints},
         year = 2025,
        month = nov,
          eid = {arXiv:2511.03787},
        pages = {arXiv:2511.03787},
          doi = {10.48550/arXiv.2511.03787},
archivePrefix = {arXiv},
       eprint = {2511.03787},
 primaryClass = {astro-ph.GA},
       adsurl = {https://ui.adsabs.harvard.edu/abs/2025arXiv251103787C}
}

@ARTICLE{casura2022,
       author = {{Casura}, Sarah and {Liske}, Jochen and {Robotham}, Aaron S.~G. and {Brough}, Sarah and {Driver}, Simon P. and {Graham}, Alister W. and {H{\"a}u{\ss}ler}, Boris and {Holwerda}, Benne W. and {Hopkins}, Andrew M. and {Kelvin}, Lee S. and {Moffett}, Amanda J. and {Taranu}, Dan S. and {Taylor}, Edward N.},
        title = "{Galaxy And Mass Assembly (GAMA): bulge-disc decomposition of KiDS data in the nearby Universe}",
      journal = {\mnras},
         year = 2022,
        month = oct,
       volume = {516},
       number = {1},
        pages = {942-974},
          doi = {10.1093/mnras/stac2267},
archivePrefix = {arXiv},
       eprint = {2208.07608},
 primaryClass = {astro-ph.GA},
       adsurl = {https://ui.adsabs.harvard.edu/abs/2022MNRAS.516..942C}
}

@ARTICLE{chabrier2003,
   author = {{Chabrier}, G.},
    title = "{Galactic Stellar and Substellar Initial Mass Function}",
  journal = {\pasp},
   eprint = {astro-ph/0304382},
     year = 2003,
    month = jul,
   volume = 115,
    pages = {763-795},
      doi = {10.1086/376392},
   adsurl = {http://adsabs.harvard.edu/abs/2003PASP..115..763C}
}

@ARTICLE{Chandrasekhar1943,
       author = {{Chandrasekhar}, S.},
        title = "{Dynamical Friction. I. General Considerations: the Coefficient of Dynamical Friction.}",
      journal = {\apj},
         year = 1943,
        month = mar,
       volume = {97},
        pages = {255},
          doi = {10.1086/144517},
       adsurl = {https://ui.adsabs.harvard.edu/abs/1943ApJ....97..255C}
}

@ARTICLE{chen2024,
       author = {{Chen}, Hongxing and {Xie}, Lizhi and {Wang}, Jie and {Hu}, Wenkai and {De Lucia}, Gabriella and {Fontanot}, Fabio and {Hirschamnn}, Michaela},
        title = "{Environmental effects on satellite galaxies from the perspective of cold gas}",
      journal = {\mnras},
         year = 2024,
        month = feb,
       volume = {528},
       number = {2},
        pages = {2451-2463},
          doi = {10.1093/mnras/stae162},
archivePrefix = {arXiv},
       eprint = {2401.07158},
 primaryClass = {astro-ph.GA},
       adsurl = {https://ui.adsabs.harvard.edu/abs/2024MNRAS.528.2451C}
}

@ARTICLE{Christodoulou1995,
       author = {{Christodoulou}, Dimitris M. and {Shlosman}, Isaac and {Tohline}, Joel E.},
        title = "{A New Criterion for Bar-forming Instability in Rapidly Rotating Gaseous and Stellar Systems. I. Axisymmetric Form}",
      journal = {\apj},
         year = 1995,
        month = apr,
       volume = {443},
        pages = {551},
          doi = {10.1086/175547},
archivePrefix = {arXiv},
       eprint = {astro-ph/9502073},
 primaryClass = {astro-ph},
       adsurl = {https://ui.adsabs.harvard.edu/abs/1995ApJ...443..551C}
}

@ARTICLE{conroy2009,
       author = {{Conroy}, Charlie and {Gunn}, James E. and {White}, Martin},
        title = "{The Propagation of Uncertainties in Stellar Population Synthesis Modeling. I. The Relevance of Uncertain Aspects of Stellar Evolution and the Initial Mass Function to the Derived Physical Properties of Galaxies}",
      journal = {\apj},
         year = 2009,
        month = jul,
       volume = {699},
       number = {1},
        pages = {486-506},
          doi = {10.1088/0004-637X/699/1/486},
archivePrefix = {arXiv},
       eprint = {0809.4261},
 primaryClass = {astro-ph},
       adsurl = {https://ui.adsabs.harvard.edu/abs/2009ApJ...699..486C}
}

@ARTICLE{covington2008,
       author = {{Covington}, M. and {Dekel}, A. and {Cox}, T.~J. and {Jonsson}, P. and {Primack}, J.~R.},
        title = "{Predicting the properties of the remnants of dissipative galaxy mergers}",
      journal = {\mnras},
         year = 2008,
        month = feb,
       volume = {384},
       number = {1},
        pages = {94-106},
          doi = {10.1111/j.1365-2966.2007.12601.x},
archivePrefix = {arXiv},
       eprint = {0710.4584},
 primaryClass = {astro-ph},
       adsurl = {https://ui.adsabs.harvard.edu/abs/2008MNRAS.384...94C}
}

@ARTICLE{Dekel&Burkert2014,
       author = {{Dekel}, A. and {Burkert}, A.},
        title = "{Wet disc contraction to galactic blue nuggets and quenching to red nuggets}",
      journal = {\mnras},
         year = 2014,
        month = feb,
       volume = {438},
       number = {2},
        pages = {1870-1879},
          doi = {10.1093/mnras/stt2331},
archivePrefix = {arXiv},
       eprint = {1310.1074},
 primaryClass = {astro-ph.CO},
       adsurl = {https://ui.adsabs.harvard.edu/abs/2014MNRAS.438.1870D}
}

@ARTICLE{delucia2007,
   author = {{De Lucia}, G. and {Blaizot}, J.},
    title = "{The hierarchical formation of the brightest cluster galaxies}",
  journal = {\mnras},
   eprint = {astro-ph/0606519},
     year = 2007,
    month = feb,
   volume = 375,
    pages = {2-14},
      doi = {10.1111/j.1365-2966.2006.11287.x},
   adsurl = {http://adsabs.harvard.edu/abs/2007MNRAS.375....2D}
}

@ARTICLE{delucia2010,
       author = {{De Lucia}, Gabriella and {Boylan-Kolchin}, Michael and
         {Benson}, Andrew J. and {Fontanot}, Fabio and {Monaco}, Pierluigi},
        title = "{A semi-analytic model comparison - gas cooling and galaxy mergers}",
      journal = {\mnras},
         year = 2010,
        month = aug,
       volume = {406},
       number = {3},
        pages = {1533-1552},
          doi = {10.1111/j.1365-2966.2010.16806.x},
archivePrefix = {arXiv},
       eprint = {1003.3021},
 primaryClass = {astro-ph.CO},
       adsurl = {https://ui.adsabs.harvard.edu/abs/2010MNRAS.406.1533D}
}

@ARTICLE{delucia2011,
       author = {{De Lucia}, Gabriella and {Fontanot}, Fabio and {Wilman}, David},
        title = "{What determines the fraction of elliptical galaxies in clusters?}",
      journal = {\mnras},
         year = 2012,
        month = jan,
       volume = {419},
       number = {2},
        pages = {1324-1330},
          doi = {10.1111/j.1365-2966.2011.19789.x},
archivePrefix = {arXiv},
       eprint = {1109.2599},
 primaryClass = {astro-ph.CO},
       adsurl = {https://ui.adsabs.harvard.edu/abs/2012MNRAS.419.1324D}
}

@ARTICLE{delucia2012,
   author = {{De Lucia}, G. and {Weinmann}, S. and {Poggianti}, B.~M. and 
	{Arag{\'o}n-Salamanca}, A. and {Zaritsky}, D.},
    title = "{The environmental history of group and cluster galaxies in a {$\Lambda$} cold dark matter universe}",
  journal = {\mnras},
archivePrefix = "arXiv",
   eprint = {1111.6590},
     year = 2012,
    month = jun,
   volume = 423,
    pages = {1277-1292},
      doi = {10.1111/j.1365-2966.2012.20983.x},
   adsurl = {http://adsabs.harvard.edu/abs/2012MNRAS.423.1277D}
}

@ARTICLE{delucia2014,
   author = {{De Lucia}, G. and {Tornatore}, L. and {Frenk}, C.~S. and {Helmi}, A. and 
	{Navarro}, J.~F. and {White}, S.~D.~M.},
    title = "{Elemental abundances in Milky Way-like galaxies from a hierarchical galaxy formation model}",
  journal = {\mnras},
archivePrefix = "arXiv",
   eprint = {1407.7867},
     year = 2014,
    month = nov,
   volume = 445,
    pages = {970-987},
      doi = {10.1093/mnras/stu1752},
   adsurl = {http://adsabs.harvard.edu/abs/2014MNRAS.445..970D}
}

@ARTICLE{delucia2024,
       author = {{De Lucia}, Gabriella and {Fontanot}, Fabio and {Xie}, Lizhi and {Hirschmann}, Michaela},
        title = "{Tracing the Quenching Journey across Cosmic Time}",
      journal = {arXiv e-prints},
         year = 2024,
        month = jan,
          eid = {arXiv:2401.06211},
        pages = {arXiv:2401.06211},
          doi = {10.48550/arXiv.2401.06211},
archivePrefix = {arXiv},
       eprint = {2401.06211},
 primaryClass = {astro-ph.GA},
       adsurl = {https://ui.adsabs.harvard.edu/abs/2024arXiv240106211D}
}

@ARTICLE{dejong2013,
       author = {{de Jong}, Jelte T.~A. and {Verdoes Kleijn}, Gijs A. and {Kuijken}, Konrad H. and {Valentijn}, Edwin A.},
        title = "{The Kilo-Degree Survey}",
      journal = {Experimental Astronomy},
         year = 2013,
        month = jan,
       volume = {35},
       number = {1-2},
        pages = {25-44},
          doi = {10.1007/s10686-012-9306-1},
archivePrefix = {arXiv},
       eprint = {1206.1254},
 primaryClass = {astro-ph.CO},
       adsurl = {https://ui.adsabs.harvard.edu/abs/2013ExA....35...25D}
}

@ARTICLE{dressler1980,
       author = {{Dressler}, A.},
        title = "{Galaxy morphology in rich clusters: implications for the formation and evolution of galaxies.}",
      journal = {\apj},
         year = 1980,
        month = mar,
       volume = {236},
        pages = {351-365},
          doi = {10.1086/157753},
       adsurl = {https://ui.adsabs.harvard.edu/abs/1980ApJ...236..351D}
}

@ARTICLE{driver2011,
       author = {{Driver}, S.~P. and {Hill}, D.~T. and {Kelvin}, L.~S. and {Robotham}, A.~S.~G. and {Liske}, J. and {Norberg}, P. and {Baldry}, I.~K. and {Bamford}, S.~P. and {Hopkins}, A.~M. and {Loveday}, J. and {Peacock}, J.~A. and {Andrae}, E. and {Bland-Hawthorn}, J. and {Brough}, S. and {Brown}, M.~J.~I. and {Cameron}, E. and {Ching}, J.~H.~Y. and {Colless}, M. and {Conselice}, C.~J. and {Croom}, S.~M. and {Cross}, N.~J.~G. and {de Propris}, R. and {Dye}, S. and {Drinkwater}, M.~J. and {Ellis}, S. and {Graham}, Alister W. and {Grootes}, M.~W. and {Gunawardhana}, M. and {Jones}, D.~H. and {van Kampen}, E. and {Maraston}, C. and {Nichol}, R.~C. and {Parkinson}, H.~R. and {Phillipps}, S. and {Pimbblet}, K. and {Popescu}, C.~C. and {Prescott}, M. and {Roseboom}, I.~G. and {Sadler}, E.~M. and {Sansom}, A.~E. and {Sharp}, R.~G. and {Smith}, D.~J.~B. and {Taylor}, E. and {Thomas}, D. and {Tuffs}, R.~J. and {Wijesinghe}, D. and {Dunne}, L. and {Frenk}, C.~S. and {Jarvis}, M.~J. and {Madore}, B.~F. and {Meyer}, M.~J. and {Seibert}, M. and {Staveley-Smith}, L. and {Sutherland}, W.~J. and {Warren}, S.~J.},
        title = "{Galaxy and Mass Assembly (GAMA): survey diagnostics and core data release}",
      journal = {\mnras},
         year = 2011,
        month = may,
       volume = {413},
       number = {2},
        pages = {971-995},
          doi = {10.1111/j.1365-2966.2010.18188.x},
archivePrefix = {arXiv},
       eprint = {1009.0614},
 primaryClass = {astro-ph.CO},
       adsurl = {https://ui.adsabs.harvard.edu/abs/2011MNRAS.413..971D}
}

@ARTICLE{elmegreen2008,
       author = {{Elmegreen}, Bruce G. and {Klessen}, Ralf S. and {Wilson}, Christine D.},
        title = "{On the Constancy of the Characteristic Mass of Young Stars}",
      journal = {\apj},
         year = 2008,
        month = jul,
       volume = {681},
       number = {1},
        pages = {365-374},
          doi = {10.1086/588725},
archivePrefix = {arXiv},
       eprint = {0803.4411},
 primaryClass = {astro-ph},
       adsurl = {https://ui.adsabs.harvard.edu/abs/2008ApJ...681..365E}
}

@ARTICLE{efstathiou1982,
       author = {{Efstathiou}, G. and {Lake}, G. and {Negroponte}, J.},
        title = "{The stability and masses of disc galaxies.}",
      journal = {\mnras},
         year = 1982,
        month = jun,
       volume = {199},
        pages = {1069-1088},
          doi = {10.1093/mnras/199.4.1069},
       adsurl = {https://ui.adsabs.harvard.edu/abs/1982MNRAS.199.1069E}
}

@ARTICLE{fontanot2015,
       author = {{Fontanot}, Fabio and {Macci{\`o}}, Andrea V. and {Hirschmann}, Michaela and {De Lucia}, Gabriella and {Kannan}, Rahul and {Somerville}, Rachel S. and {Wilman}, Dave},
        title = "{On the dependence of galaxy morphologies on galaxy mergers}",
      journal = {\mnras},
         year = 2015,
        month = aug,
       volume = {451},
       number = {3},
        pages = {2968-2977},
          doi = {10.1093/mnras/stv1119},
archivePrefix = {arXiv},
       eprint = {1507.04748},
 primaryClass = {astro-ph.GA},
       adsurl = {https://ui.adsabs.harvard.edu/abs/2015MNRAS.451.2968F}
}

@ARTICLE{fontanot2020,
       author = {{Fontanot}, Fabio and {De Lucia}, Gabriella and {Hirschmann}, Michaela and {Xie}, Lizhi and {Monaco}, Pierluigi and {Menci}, Nicola and {Fiore}, Fabrizio and {Feruglio}, Chiara and {Cristiani}, Stefano and {Shankar}, Francesco},
        title = "{The rise of active galactic nuclei in the galaxy evolution and assembly semi-analytic model}",
      journal = {\mnras},
         year = 2020,
        month = aug,
       volume = {496},
       number = {3},
        pages = {3943-3960},
          doi = {10.1093/mnras/staa1716},
archivePrefix = {arXiv},
       eprint = {2002.10576},
 primaryClass = {astro-ph.CO},
       adsurl = {https://ui.adsabs.harvard.edu/abs/2020MNRAS.496.3943F}
}

@ARTICLE{fontanot2021,
       author = {{Fontanot}, Fabio and {Calabr{\`o}}, Antonello and {Talia}, Margherita and {Mannucci}, Filippo and {Castellano}, Marco and {Cresci}, Giovanni and {De Lucia}, Gabriella and {Gallazzi}, Anna and {Hirschmann}, Michaela and {Pentericci}, Laura and {Xie}, Lizhi and {Amorin}, Ricardo and {Bolzonella}, Micol and {Bongiorno}, Angela and {Cucciati}, Olga and {Cullen}, Fergus and {Fynbo}, Johan P.~U. and {Hathi}, Nimish and {Hibon}, Pascale and {McLure}, Ross J. and {Pozzetti}, Lucia},
        title = "{The evolution of the mass-metallicity relations from the VANDELS survey and the GAEA semi-analytic model}",
      journal = {\mnras},
         year = 2021,
        month = jul,
       volume = {504},
       number = {3},
        pages = {4481-4492},
          doi = {10.1093/mnras/stab1213},
archivePrefix = {arXiv},
       eprint = {2104.08295},
 primaryClass = {astro-ph.GA},
       adsurl = {https://ui.adsabs.harvard.edu/abs/2021MNRAS.504.4481F}
}

@ARTICLE{fontanot2025,
       author = {{Fontanot}, Fabio and {De Lucia}, Gabriella and {Xie}, Lizhi and {Hirschmann}, Michaela and {Baugh}, Carlton and {Helly}, John C.},
        title = "{Galaxy assembly and evolution in the P-Millennium simulation: Galaxy clustering}",
      journal = {\aap},
         year = 2025,
        month = jul,
       volume = {699},
          eid = {A108},
        pages = {A108},
          doi = {10.1051/0004-6361/202452029},
archivePrefix = {arXiv},
       eprint = {2409.02194},
 primaryClass = {astro-ph.GA},
       adsurl = {https://ui.adsabs.harvard.edu/abs/2025A&A...699A.108F}
}

@ARTICLE{Genin2025,
       author = {{Genin}, Aur{\'e}lien and {Shuntov}, Marko and {Brammer}, Gabe and {Allen}, Natalie and {Ito}, Kei and {Magdis}, Georgios and {Matharu}, Jasleen and {Oesch}, Pascal A. and {Toft}, Sune and {Valentino}, Francesco},
        title = "{DAWN JWST Archive: Morphology from profile fitting of over 340 000 galaxies in major JWST fields: Morphology evolution with redshift and galaxy type}",
      journal = {\aap},
         year = 2025,
        month = jul,
       volume = {699},
          eid = {A343},
        pages = {A343},
          doi = {10.1051/0004-6361/202555504},
archivePrefix = {arXiv},
       eprint = {2505.21622},
 primaryClass = {astro-ph.GA},
       adsurl = {https://ui.adsabs.harvard.edu/abs/2025A&A...699A.343G}
}

@ARTICLE{guo2011,
       author = {{Guo}, Qi and {White}, Simon and {Boylan-Kolchin}, Michael and {De Lucia}, Gabriella and {Kauffmann}, Guinevere and {Lemson}, Gerard and {Li}, Cheng and {Springel}, Volker and {Weinmann}, Simone},
        title = "{From dwarf spheroidals to cD galaxies: simulating the galaxy population in a {\ensuremath{\Lambda}}CDM cosmology}",
      journal = {\mnras},
         year = 2011,
        month = may,
       volume = {413},
       number = {1},
        pages = {101-131},
          doi = {10.1111/j.1365-2966.2010.18114.x},
archivePrefix = {arXiv},
       eprint = {1006.0106},
 primaryClass = {astro-ph.CO},
       adsurl = {https://ui.adsabs.harvard.edu/abs/2011MNRAS.413..101G}
}

@ARTICLE{hirschmann2014,
       author = {{Hirschmann}, Michaela and {De Lucia}, Gabriella and {Wilman}, Dave and {Weinmann}, Simone and {Iovino}, Angela and {Cucciati}, Olga and {Zibetti}, Stefano and {Villalobos}, {\'A}lvaro},
        title = "{The influence of the environmental history on quenching star formation in a {\ensuremath{\Lambda}} cold dark matter universe}",
      journal = {\mnras},
         year = 2014,
        month = nov,
       volume = {444},
       number = {3},
        pages = {2938-2959},
          doi = {10.1093/mnras/stu1609},
archivePrefix = {arXiv},
       eprint = {1407.5621},
 primaryClass = {astro-ph.GA},
       adsurl = {https://ui.adsabs.harvard.edu/abs/2014MNRAS.444.2938H}
}

@ARTICLE{hirschmann2016,
   author = {{Hirschmann}, M. and {De Lucia}, G. and {Fontanot}, F.},
    title = "{Galaxy assembly, stellar feedback and metal enrichment: the view from the GAEA model}",
  journal = {\mnras},
archivePrefix = "arXiv",
   eprint = {1512.04531},
     year = 2016,
    month = sep,
   volume = 461,
    pages = {1760-1785},
      doi = {10.1093/mnras/stw1318},
   adsurl = {http://adsabs.harvard.edu/abs/2016MNRAS.461.1760H}
}

@ARTICLE{hubble1926,
       author = {{Hubble}, E.~P.},
        title = "{Extragalactic nebulae.}",
      journal = {\apj},
         year = 1926,
        month = dec,
       volume = {64},
        pages = {321-369},
          doi = {10.1086/143018},
       adsurl = {https://ui.adsabs.harvard.edu/abs/1926ApJ....64..321H}
}

@ARTICLE{huertas-company2024,
       author = {{Huertas-Company}, M. and {Iyer}, K.~G. and {Angeloudi}, E. and {Bagley}, M.~B. and {Finkelstein}, S.~L. and {Kartaltepe}, J. and {McGrath}, E.~J. and {Sarmiento}, R. and {Vega-Ferrero}, J. and {Arrabal Haro}, P. and {Behroozi}, P. and {Buitrago}, F. and {Cheng}, Y. and {Costantin}, L. and {Dekel}, A. and {Dickinson}, M. and {Elbaz}, D. and {Grogin}, N.~A. and {Hathi}, N.~P. and {Holwerda}, B.~W. and {Koekemoer}, A.~M. and {Lucas}, R.~A. and {Papovich}, C. and {P{\'e}rez-Gonz{\'a}lez}, P.~G. and {Pirzkal}, N. and {Seill{\'e}}, L.-M. and {de la Vega}, A. and {Wuyts}, S. and {Yang}, G. and {Yung}, L.~Y.~A.},
        title = "{Galaxy morphology from z {\ensuremath{\sim}} 6 through the lens of JWST}",
      journal = {\aap},
         year = 2024,
        month = may,
       volume = {685},
          eid = {A48},
        pages = {A48},
          doi = {10.1051/0004-6361/202346800},
archivePrefix = {arXiv},
       eprint = {2305.02478},
 primaryClass = {astro-ph.GA},
       adsurl = {https://ui.adsabs.harvard.edu/abs/2024A&A...685A..48H}
}

@ARTICLE{huertas-company2025,
       author = {{Huertas-Company}, M. and {Shuntov}, M. and {Dong}, Y. and {Walmsley}, M. and {Ilbert}, O. and {McCracken}, H.~J. and {Akins}, H.~B. and {Allen}, N. and {Casey}, C.~M. and {Costantin}, L. and {Daddi}, E. and {Dekel}, A. and {Franco}, M. and {Garland}, I.~L. and {G{\'e}ron}, T. and {Gozaliasl}, G. and {Hirschmann}, M. and {Kartaltepe}, J.~S. and {Koekemoer}, A.~M. and {Lintott}, C. and {Liu}, D. and {Lucas}, R. and {Masters}, K. and {Pacucci}, F. and {Paquereau}, L. and {P{\'e}rez-Gonz{\'a}lez}, P.~G. and {Rhodes}, J.~D. and {Robertson}, B.~E. and {Simmons}, B. and {Smethurst}, R. and {Toft}, S. and {Yang}, L.},
        title = "{COSMOS-Web: The emergence of the Hubble sequence}",
      journal = {\aap},
         year = 2025,
        month = dec,
       volume = {704},
          eid = {A94},
        pages = {A94},
          doi = {10.1051/0004-6361/202553782},
archivePrefix = {arXiv},
       eprint = {2502.03532},
 primaryClass = {astro-ph.GA},
       adsurl = {https://ui.adsabs.harvard.edu/abs/2025A&A...704A..94H}
}

@ARTICLE{hopkins2009,
       author = {{Hopkins}, Philip F. and {Somerville}, Rachel S. and {Cox}, Thomas J. and {Hernquist}, Lars and {Jogee}, Shardha and {Kere{\v{s}}}, Dusan and {Ma}, Chung-Pei and {Robertson}, Brant and {Stewart}, Kyle},
        title = "{The effects of gas on morphological transformation in mergers: implications for bulge and disc demographics}",
      journal = {\mnras},
         year = 2009,
        month = aug,
       volume = {397},
       number = {2},
        pages = {802-814},
          doi = {10.1111/j.1365-2966.2009.14983.x},
archivePrefix = {arXiv},
       eprint = {0901.4111},
 primaryClass = {astro-ph.CO},
       adsurl = {https://ui.adsabs.harvard.edu/abs/2009MNRAS.397..802H}
}

@ARTICLE{hopkins2010,
       author = {{Hopkins}, Philip F. and {Bundy}, Kevin and {Croton}, Darren and {Hernquist}, Lars and {Keres}, Dusan and {Khochfar}, Sadegh and {Stewart}, Kyle and {Wetzel}, Andrew and {Younger}, Joshua D.},
        title = "{Mergers and Bulge Formation in {\ensuremath{\Lambda}}CDM: Which Mergers Matter?}",
      journal = {\apj},
         year = 2010,
        month = may,
       volume = {715},
       number = {1},
        pages = {202-229},
          doi = {10.1088/0004-637X/715/1/202},
archivePrefix = {arXiv},
       eprint = {0906.5357},
 primaryClass = {astro-ph.CO},
       adsurl = {https://ui.adsabs.harvard.edu/abs/2010ApJ...715..202H}
}

@ARTICLE{Irodotou2019,
       author = {{Irodotou}, Dimitrios and {Thomas}, Peter A. and {Henriques}, Bruno M. and {Sargent}, Mark T. and {Hislop}, Jessica M.},
        title = "{Morphological evolution and galactic sizes in the L-Galaxies SA model}",
      journal = {\mnras},
         year = 2019,
        month = nov,
       volume = {489},
       number = {3},
        pages = {3609-3624},
          doi = {10.1093/mnras/stz2365},
archivePrefix = {arXiv},
       eprint = {1810.05173},
 primaryClass = {astro-ph.GA},
       adsurl = {https://ui.adsabs.harvard.edu/abs/2019MNRAS.489.3609I}
}

@ARTICLE{jaffe1983,
   author = {{Jaffe}, W.},
    title = "{A simple model for the distribution of light in spherical galaxies}",
  journal = {\mnras},
     year = 1983,
    month = mar,
   volume = 202,
    pages = {995-999},
      doi = {10.1093/mnras/202.4.995},
   adsurl = {http://adsabs.harvard.edu/abs/1983MNRAS.202..995J}
}

@ARTICLE{jiang2010,
       author = {{Jiang}, C.~Y. and {Jing}, Y.~P. and {Lin}, W.~P.},
        title = "{Influence of baryonic physics on the merger timescale of galaxies in N-body/hydrodynamical simulations}",
      journal = {\aap},
         year = 2010,
        month = feb,
       volume = {510},
          eid = {A60},
        pages = {A60},
          doi = {10.1051/0004-6361/200913257},
archivePrefix = {arXiv},
       eprint = {0902.3734},
 primaryClass = {astro-ph.CO},
       adsurl = {https://ui.adsabs.harvard.edu/abs/2010A&A...510A..60J}
}

@ARTICLE{kartaltepe2023,
       author = {{Kartaltepe}, Jeyhan S. and {Rose}, Caitlin and {Vanderhoof}, Brittany N. and {McGrath}, Elizabeth J. and {Costantin}, Luca and {Cox}, Isabella G. and {Yung}, L.~Y. Aaron and {Kocevski}, Dale D. and {Wuyts}, Stijn and {Ferguson}, Henry C. and {Bagley}, Micaela B. and {Finkelstein}, Steven L. and {Amor{\'\i}n}, Ricardo O. and {Andrews}, Brett H. and {Arrabal Haro}, Pablo and {Backhaus}, Bren E. and {Behroozi}, Peter and {Bisigello}, Laura and {Calabr{\`o}}, Antonello and {Casey}, Caitlin M. and {Coogan}, Rosemary T. and {Cooper}, M.~C. and {Croton}, Darren and {de la Vega}, Alexander and {Dickinson}, Mark and {Fontana}, Adriano and {Franco}, Maximilien and {Grazian}, Andrea and {Grogin}, Norman A. and {Hathi}, Nimish P. and {Holwerda}, Benne W. and {Huertas-Company}, Marc and {Iyer}, Kartheik G. and {Jogee}, Shardha and {Jung}, Intae and {Kewley}, Lisa J. and {Kirkpatrick}, Allison and {Koekemoer}, Anton M. and {Liu}, James and {Lotz}, Jennifer M. and {Lucas}, Ray A. and {Newman}, Jeffrey A. and {Pacifici}, Camilla and {Pandya}, Viraj and {Papovich}, Casey and {Pentericci}, Laura and {P{\'e}rez-Gonz{\'a}lez}, Pablo G. and {Petersen}, Jayse and {Pirzkal}, Nor and {Rafelski}, Marc and {Ravindranath}, Swara and {Simons}, Raymond C. and {Snyder}, Gregory F. and {Somerville}, Rachel S. and {Stanway}, Elizabeth R. and {Straughn}, Amber N. and {Tacchella}, Sandro and {Trump}, Jonathan R. and {Vega-Ferrero}, Jes{\'u}s and {Wilkins}, Stephen M. and {Yang}, Guang and {Zavala}, Jorge A.},
        title = "{CEERS Key Paper. III. The Diversity of Galaxy Structure and Morphology at z = 3-9 with JWST}",
      journal = {\apjl},
         year = 2023,
        month = mar,
       volume = {946},
       number = {1},
          eid = {L15},
        pages = {L15},
          doi = {10.3847/2041-8213/acad01},
archivePrefix = {arXiv},
       eprint = {2210.14713},
 primaryClass = {astro-ph.GA},
       adsurl = {https://ui.adsabs.harvard.edu/abs/2023ApJ...946L..15K}
}

@ARTICLE{kauffmann1993,
       author = {{Kauffmann}, G. and {White}, S.~D.~M. and {Guiderdoni}, B.},
        title = "{The formation and evolution of galaxies within merging dark matter haloes.}",
      journal = {\mnras},
         year = 1993,
        month = sep,
       volume = {264},
        pages = {201-218},
          doi = {10.1093/mnras/264.1.201},
       adsurl = {https://ui.adsabs.harvard.edu/abs/1993MNRAS.264..201K}
}

@ARTICLE{kannan2015,
       author = {{Kannan}, Rahul and {Macci{\`o}}, Andrea V. and {Fontanot}, Fabio and {Moster}, Benjamin P. and {Karman}, Wouter and {Somerville}, Rachel S.},
        title = "{From discs to bulges: effect of mergers on the morphology of galaxies}",
      journal = {\mnras},
         year = 2015,
        month = oct,
       volume = {452},
       number = {4},
        pages = {4347-4360},
          doi = {10.1093/mnras/stv1633},
archivePrefix = {arXiv},
       eprint = {1507.04746},
 primaryClass = {astro-ph.GA},
       adsurl = {https://ui.adsabs.harvard.edu/abs/2015MNRAS.452.4347K}
}

@ARTICLE{kawinwanichakij2026,
       author = {{Kawinwanichakij}, Lalitwadee and {Glazebrook}, Karl and {Nanayakkara}, Themiya and {Kacprzak}, Glenn G. and {Chittenden}, Harry George and {Jacobs}, Colin and {Chandro-G{\'o}mez}, {\'A}ngel and {Lagos}, Claudia and {Marchesini}, Danilo and {Mart{\`\i}nez-Mar{\`\i}n}, M. and {Oesch}, Pascal A. and {Remus}, Rhea-Silvia},
        title = "{Connecting Environment, Star Formation History, and Morphology of Massive Quiescent Galaxies at 3 < z < 4 with JWST}",
      journal = {\apj},
         year = 2026,
        month = jan,
       volume = {997},
       number = {1},
          eid = {29},
        pages = {29},
          doi = {10.3847/1538-4357/ae0a18},
archivePrefix = {arXiv},
       eprint = {2505.03089},
 primaryClass = {astro-ph.GA},
       adsurl = {https://ui.adsabs.harvard.edu/abs/2026ApJ...997...29K}
}

@INPROCEEDINGS{kormendy2016,
       author = {{Kormendy}, John},
        title = "{Elliptical Galaxies and Bulges of Disc Galaxies: Summary of Progress and Outstanding Issues}",
    booktitle = {Galactic Bulges},
         year = 2016,
       editor = {{Laurikainen}, Eija and {Peletier}, Reynier and {Gadotti}, Dimitri},
       series = {Astrophysics and Space Science Library},
       volume = {418},
        month = jan,
        pages = {431},
          doi = {10.1007/978-3-319-19378-6_16},
archivePrefix = {arXiv},
       eprint = {1504.03330},
 primaryClass = {astro-ph.GA},
       adsurl = {https://ui.adsabs.harvard.edu/abs/2016ASSL..418..431K}
}

@ARTICLE{lehnert2014,
       author = {{Lehnert}, Matthew D. and {Di Matteo}, Paola and {Haywood}, Misha and {Snaith}, Owain N.},
        title = "{The Milky Way as a High-redshift Galaxy: The Importance of Thick Disk Formation in Galaxies}",
      journal = {\apjl},
         year = 2014,
        month = jul,
       volume = {789},
       number = {2},
          eid = {L30},
        pages = {L30},
          doi = {10.1088/2041-8205/789/2/L30},
archivePrefix = {arXiv},
       eprint = {1406.3039},
 primaryClass = {astro-ph.GA},
       adsurl = {https://ui.adsabs.harvard.edu/abs/2014ApJ...789L..30L}
}

@ARTICLE{li2009,
       author = {{Li}, Cheng and {White}, Simon D.~M.},
        title = "{The distribution of stellar mass in the low-redshift Universe}",
      journal = {\mnras},
         year = 2009,
        month = oct,
       volume = {398},
       number = {4},
        pages = {2177-2187},
          doi = {10.1111/j.1365-2966.2009.15268.x},
archivePrefix = {arXiv},
       eprint = {0901.0706},
 primaryClass = {astro-ph.CO},
       adsurl = {https://ui.adsabs.harvard.edu/abs/2009MNRAS.398.2177L}
}

@ARTICLE{liu2021,
       author = {{Liu}, Shuang and {Gu}, Yizhou and {Yuan}, Qirong and {Lu}, Shiying and {Bao}, Min and {Fang}, Guanwen and {Fan}, Lulu},
        title = "{Morphological Transformation and Star Formation Quenching of Massive Galaxies at 0.5 {\ensuremath{\leq}} z {\ensuremath{\leq}} 2.5 in 3D-HST/CANDELS}",
      journal = {\apj},
         year = 2021,
        month = dec,
       volume = {923},
       number = {1},
          eid = {46},
        pages = {46},
          doi = {10.3847/1538-4357/ac2817},
archivePrefix = {arXiv},
       eprint = {2110.12704},
 primaryClass = {astro-ph.GA},
       adsurl = {https://ui.adsabs.harvard.edu/abs/2021ApJ...923...46L}
}

@ARTICLE{lotz2008,
       author = {{Lotz}, Jennifer M. and {Davis}, M. and {Faber}, S.~M. and {Guhathakurta}, P. and {Gwyn}, S. and {Huang}, J. and {Koo}, D.~C. and {Le Floc'h}, E. and {Lin}, Lihwai and {Newman}, J. and et al.},
        title = "{The Evolution of Galaxy Mergers and Morphology at z < 1.2 in the Extended Groth Strip}",
      journal = {\apj},
         year = 2008,
        month = jan,
       volume = {672},
       number = {1},
        pages = {177-197},
          doi = {10.1086/523659},
archivePrefix = {arXiv},
       eprint = {astro-ph/0602088},
 primaryClass = {astro-ph},
       adsurl = {https://ui.adsabs.harvard.edu/abs/2008ApJ...672..177L}
}

@ARTICLE{ludlow2023,
       author = {{Ludlow}, Aaron D. and {Fall}, S. Michael and {Wilkinson}, Matthew J. and {Schaye}, Joop and {Obreschkow}, Danail},
        title = "{Spurious heating of stellar motions by dark matter particles in cosmological simulations of galaxy formation}",
      journal = {\mnras},
         year = 2023,
        month = nov,
       volume = {525},
       number = {4},
        pages = {5614-5630},
          doi = {10.1093/mnras/stad2615},
archivePrefix = {arXiv},
       eprint = {2306.05753},
 primaryClass = {astro-ph.GA},
       adsurl = {https://ui.adsabs.harvard.edu/abs/2023MNRAS.525.5614L}
}

@ARTICLE{martin2018,
       author = {{Martin}, G. and {Kaviraj}, S. and {Devriendt}, J.~E.~G. and {Dubois}, Y. and {Pichon}, C.},
        title = "{The role of mergers in driving morphological transformation over cosmic time}",
      journal = {\mnras},
         year = 2018,
        month = oct,
       volume = {480},
       number = {2},
        pages = {2266-2283},
          doi = {10.1093/mnras/sty1936},
archivePrefix = {arXiv},
       eprint = {1807.08761},
 primaryClass = {astro-ph.GA},
       adsurl = {https://ui.adsabs.harvard.edu/abs/2018MNRAS.480.2266M}
}

@ARTICLE{mendel2014,
       author = {{Mendel}, J. Trevor and {Simard}, Luc and {Palmer}, Michael and {Ellison}, Sara L. and {Patton}, David R.},
        title = "{A Catalog of Bulge, Disk, and Total Stellar Mass Estimates for the Sloan Digital Sky Survey}",
      journal = {\apjs},
         year = 2014,
        month = jan,
       volume = {210},
       number = {1},
          eid = {3},
        pages = {3},
          doi = {10.1088/0067-0049/210/1/3},
archivePrefix = {arXiv},
       eprint = {1310.8304},
 primaryClass = {astro-ph.CO},
       adsurl = {https://ui.adsabs.harvard.edu/abs/2014ApJS..210....3M}
}

@ARTICLE{mihos1996,
       author = {{Mihos}, J. Christopher and {Hernquist}, Lars},
        title = "{Gasdynamics and Starbursts in Major Mergers}",
      journal = {\apj},
         year = 1996,
        month = jun,
       volume = {464},
        pages = {641},
          doi = {10.1086/177353},
archivePrefix = {arXiv},
       eprint = {astro-ph/9512099},
 primaryClass = {astro-ph},
       adsurl = {https://ui.adsabs.harvard.edu/abs/1996ApJ...464..641M}
}

@ARTICLE{moffett2016,
       author = {{Moffett}, Amanda J. and {Ingarfield}, Stephen A. and {Driver}, Simon P. and {Robotham}, Aaron S.~G. and {Kelvin}, Lee S. and {Lange}, Rebecca and {Me{\v{s}}tri{\'c}}, Uro{\v{s}} and {Alpaslan}, Mehmet and {Baldry}, Ivan K. and {Bland-Hawthorn}, Joss and {Brough}, Sarah and {Cluver}, Michelle E. and {Davies}, Luke J.~M. and {Holwerda}, Benne W. and {Hopkins}, Andrew M. and {Kafle}, Prajwal R. and {Kennedy}, Rebecca and {Norberg}, Peder and {Taylor}, Edward N.},
        title = "{Galaxy And Mass Assembly (GAMA): the stellar mass budget by galaxy type}",
      journal = {\mnras},
         year = 2016,
        month = apr,
       volume = {457},
       number = {2},
        pages = {1308-1319},
          doi = {10.1093/mnras/stv2883},
archivePrefix = {arXiv},
       eprint = {1512.02342},
 primaryClass = {astro-ph.GA},
       adsurl = {https://ui.adsabs.harvard.edu/abs/2016MNRAS.457.1308M}
}

@ARTICLE{moster2011,
       author = {{Moster}, Benjamin P. and {Macci{\`o}}, Andrea V. and {Somerville}, Rachel S. and {Naab}, Thorsten and {Cox}, T.~J.},
        title = "{The effects of a hot gaseous halo in galaxy major mergers}",
      journal = {\mnras},
         year = 2011,
        month = aug,
       volume = {415},
       number = {4},
        pages = {3750-3770},
          doi = {10.1111/j.1365-2966.2011.18984.x},
archivePrefix = {arXiv},
       eprint = {1104.0246},
 primaryClass = {astro-ph.GA},
       adsurl = {https://ui.adsabs.harvard.edu/abs/2011MNRAS.415.3750M}
}

@ARTICLE{osman2026,
       author = {{Osman}, Omima and {De Lucia}, Gabriella and {Fontanot}, Fabio and {Xie}, Lizhi and {Hirschmann}, Michaela},
        title = "{Dust evolution across cosmic times as seen through DUSTY-GAEA}",
      journal = {\aap},
         year = 2026,
        month = jun,
       volume = {710},
          eid = {A306},
        pages = {A306},
          doi = {10.1051/0004-6361/202558572},
       adsurl = {https://ui.adsabs.harvard.edu/abs/2026A&A...710A.306O}
}

@ARTICLE{pei2024,
       author = {{Pei}, Wenxiang and {Guo}, Qi and {Li}, Ming and {Wang}, Qiao and {Han}, Jiaxin and {Hu}, Jia and {Su}, Tong and {Gao}, Liang and {Wang}, Jie and {Luo}, Yu and et al.},
        title = "{Simulating emission line galaxies for the next generation of large-scale structure surveys}",
      journal = {\mnras},
         year = 2024,
        month = apr,
       volume = {529},
       number = {4},
        pages = {4958-4979},
          doi = {10.1093/mnras/stae866},
archivePrefix = {arXiv},
       eprint = {2404.00092},
 primaryClass = {astro-ph.GA},
       adsurl = {https://ui.adsabs.harvard.edu/abs/2024MNRAS.529.4958P}
}

@ARTICLE{pfeffer2023,
       author = {{Pfeffer}, Joel and {Cavanagh}, Mitchell K. and {Bekki}, Kenji and {Couch}, Warrick J. and {Drinkwater}, Michael J. and {Forbes}, Duncan A. and {Koribalski}, B{\"a}rbel S.},
        title = "{The galaxy morphology-density relation in the EAGLE simulation}",
      journal = {\mnras},
         year = 2023,
        month = feb,
       volume = {518},
       number = {4},
        pages = {5260-5278},
          doi = {10.1093/mnras/stac3466},
archivePrefix = {arXiv},
       eprint = {2212.08748},
 primaryClass = {astro-ph.GA},
       adsurl = {https://ui.adsabs.harvard.edu/abs/2023MNRAS.518.5260P}
}

@ARTICLE{Pu2026,
       author = {{Pu}, Yahan and {Wang}, Lan and {Zeng}, Guangquan and {Xie}, Lizhi},
        title = "{Comparing Galaxy Merger Orbits in Hydrodynamical Simulation and in Dark Matter only Simulation}",
      journal = {Research in Astronomy and Astrophysics},
         year = 2026,
        month = nov,
       volume = {26},
       number = {11},
          eid = {115009},
        pages = {115009},
          doi = {10.1088/1674-4527/ae6cbe},
archivePrefix = {arXiv},
       eprint = {2511.17060},
 primaryClass = {astro-ph.GA},
       adsurl = {https://ui.adsabs.harvard.edu/abs/2026RAA....26k5009P}
}

@ARTICLE{rodriguez-gomez2017,
       author = {{Rodriguez-Gomez}, Vicente and {Sales}, Laura V. and {Genel}, Shy and {Pillepich}, Annalisa and {Zjupa}, Jolanta and {Nelson}, Dylan and {Griffen}, Brendan and {Torrey}, Paul and {Snyder}, Gregory F. and {Vogelsberger}, Mark and et al.},
        title = "{The role of mergers and halo spin in shaping galaxy morphology}",
      journal = {\mnras},
         year = 2017,
        month = may,
       volume = {467},
       number = {3},
        pages = {3083-3098},
          doi = {10.1093/mnras/stx305},
archivePrefix = {arXiv},
       eprint = {1609.09498},
 primaryClass = {astro-ph.GA},
       adsurl = {https://ui.adsabs.harvard.edu/abs/2017MNRAS.467.3083R}
}

@ARTICLE{Scannapieco2008,
       author = {{Scannapieco}, Cecilia and {Tissera}, Patricia B. and {White}, Simon D.~M. and {Springel}, Volker},
        title = "{Effects of supernova feedback on the formation of galaxy discs}",
      journal = {\mnras},
         year = 2008,
        month = sep,
       volume = {389},
       number = {3},
        pages = {1137-1149},
          doi = {10.1111/j.1365-2966.2008.13678.x},
archivePrefix = {arXiv},
       eprint = {0804.3795},
 primaryClass = {astro-ph},
       adsurl = {https://ui.adsabs.harvard.edu/abs/2008MNRAS.389.1137S}
}

@ARTICLE{scannapieco2009,
       author = {{Scannapieco}, Cecilia and {White}, Simon D.~M. and {Springel}, Volker and {Tissera}, Patricia B.},
        title = "{The formation and survival of discs in a {\ensuremath{\Lambda}}CDM universe}",
      journal = {\mnras},
         year = 2009,
        month = jun,
       volume = {396},
       number = {2},
        pages = {696-708},
          doi = {10.1111/j.1365-2966.2009.14764.x},
archivePrefix = {arXiv},
       eprint = {0812.0976},
 primaryClass = {astro-ph},
       adsurl = {https://ui.adsabs.harvard.edu/abs/2009MNRAS.396..696S}
}

@ARTICLE{schreiber2020,
       author = {{F{\"o}rster Schreiber}, Natascha M. and {Wuyts}, Stijn},
        title = "{Star-Forming Galaxies at Cosmic Noon}",
      journal = {\araa},
         year = 2020,
        month = aug,
       volume = {58},
        pages = {661-725},
          doi = {10.1146/annurev-astro-032620-021910},
archivePrefix = {arXiv},
       eprint = {2010.10171},
 primaryClass = {astro-ph.GA},
       adsurl = {https://ui.adsabs.harvard.edu/abs/2020ARA&A..58..661F}
}

@ARTICLE{sellwood2001,
       author = {{Sellwood}, J.~A. and {Evans}, N.~W.},
        title = "{The Stability of Disks in Cusped Potentials}",
      journal = {\apj},
         year = 2001,
        month = jan,
       volume = {546},
       number = {1},
        pages = {176-188},
          doi = {10.1086/318228},
archivePrefix = {arXiv},
       eprint = {astro-ph/0006198},
 primaryClass = {astro-ph},
       adsurl = {https://ui.adsabs.harvard.edu/abs/2001ApJ...546..176S}
}

@ARTICLE{shuntov2025,
       author = {{Shuntov}, Marko and {Akins}, Hollis B. and {Paquereau}, Louise and {Casey}, Caitlin M. and {Ilbert}, Olivier and {Arango-Toro}, Rafael C. and {McCracken}, Henry Joy and {Franco}, Maximilien and {Harish}, Santosh and {Kartaltepe}, Jeyhan S. and et al.},
        title = "{COSMOS2025: The COSMOS-Web galaxy catalog of photometry, morphology, redshifts, and physical parameters from JWST, HST, and ground-based imaging}",
      journal = {\aap},
         year = 2025,
        month = dec,
       volume = {704},
          eid = {A339},
        pages = {A339},
          doi = {10.1051/0004-6361/202555799},
archivePrefix = {arXiv},
       eprint = {2506.03243},
 primaryClass = {astro-ph.GA},
       adsurl = {https://ui.adsabs.harvard.edu/abs/2025A&A...704A.339S}
}

@ARTICLE{shuntov2026,
       author = {{Shuntov}, Marko and {Ilbert}, Olivier and {Lagos}, Claudia del P. and {Toft}, Sune and {Valentino}, Francesco and {Mercier}, Wilfried and {Akins}, Hollis B. and {Binh}, Nguyen and {Brinch}, Malte and {Casey}, Caitlin M. and et al.},
        title = "{The stellar mass function of quiescent and star-forming galaxies and its dependence on morphology in COSMOS-Web}",
      journal = {\aap},
         year = 2026,
        month = mar,
       volume = {707},
          eid = {A391},
        pages = {A391},
          doi = {10.1051/0004-6361/202558022},
archivePrefix = {arXiv},
       eprint = {2511.05259},
 primaryClass = {astro-ph.GA},
       adsurl = {https://ui.adsabs.harvard.edu/abs/2026A&A...707A.391S}
}

@ARTICLE{Simard2011,
       author = {{Simard}, Luc and {Mendel}, J. Trevor and {Patton}, David R. and {Ellison}, Sara L. and {McConnachie}, Alan W.},
        title = "{A Catalog of Bulge+disk Decompositions and Updated Photometry for 1.12 Million Galaxies in the Sloan Digital Sky Survey}",
      journal = {\apjs},
         year = 2011,
        month = sep,
       volume = {196},
       number = {1},
          eid = {11},
        pages = {11},
          doi = {10.1088/0067-0049/196/1/11},
archivePrefix = {arXiv},
       eprint = {1107.1518},
 primaryClass = {astro-ph.CO},
       adsurl = {https://ui.adsabs.harvard.edu/abs/2011ApJS..196...11S}
}

@ARTICLE{spergel2003,
       author = {{Spergel}, D.~N. and {Verde}, L. and {Peiris}, H.~V. and {Komatsu}, E. and {Nolta}, M.~R. and {Bennett}, C.~L. and {Halpern}, M. and {Hinshaw}, G. and {Jarosik}, N. and {Kogut}, A. and {Limon}, M. and {Meyer}, S.~S. and {Page}, L. and {Tucker}, G.~S. and {Weiland}, J.~L. and {Wollack}, E. and {Wright}, E.~L.},
        title = "{First-Year Wilkinson Microwave Anisotropy Probe (WMAP) Observations: Determination of Cosmological Parameters}",
      journal = {\apjs},
         year = 2003,
        month = sep,
       volume = {148},
       number = {1},
        pages = {175-194},
          doi = {10.1086/377226},
archivePrefix = {arXiv},
       eprint = {astro-ph/0302209},
 primaryClass = {astro-ph},
       adsurl = {https://ui.adsabs.harvard.edu/abs/2003ApJS..148..175S}
}

@ARTICLE{springel2005,
       author = {{Springel}, Volker and {White}, Simon D.~M. and {Jenkins}, Adrian and
         {Frenk}, Carlos S. and {Yoshida}, Naoki and {Gao}, Liang and
         {Navarro}, Julio and {Thacker}, Robert and {Croton}, Darren and
         {Helly}, John and {Peacock}, John A. and {Cole}, Shaun and
         {Thomas}, Peter and {Couchman}, Hugh and {Evrard}, August and
         {Colberg}, J{\"o}rg and {Pearce}, Frazer},
        title = "{Simulations of the formation, evolution and clustering of galaxies and quasars}",
      journal = {\nat},
         year = "2005",
        month = "Jun",
       volume = {435},
       number = {7042},
        pages = {629-636},
          doi = {10.1038/nature03597},
archivePrefix = {arXiv},
       eprint = {astro-ph/0504097},
 primaryClass = {astro-ph},
       adsurl = {https://ui.adsabs.harvard.edu/abs/2005Natur.435..629S}
}

@ARTICLE{somerville2001,
       author = {{Somerville}, Rachel S. and {Primack}, Joel R. and {Faber}, S.~M.},
        title = "{The nature of high-redshift galaxies}",
      journal = {\mnras},
         year = 2001,
        month = feb,
       volume = {320},
       number = {4},
        pages = {504-528},
          doi = {10.1046/j.1365-8711.2001.03975.x},
archivePrefix = {arXiv},
       eprint = {astro-ph/9806228},
 primaryClass = {astro-ph},
       adsurl = {https://ui.adsabs.harvard.edu/abs/2001MNRAS.320..504S}
}

@ARTICLE{somerville2008,
       author = {{Somerville}, Rachel S. and {Hopkins}, Philip F. and {Cox}, Thomas J. and {Robertson}, Brant E. and {Hernquist}, Lars},
        title = "{A semi-analytic model for the co-evolution of galaxies, black holes and active galactic nuclei}",
      journal = {\mnras},
         year = 2008,
        month = dec,
       volume = {391},
       number = {2},
        pages = {481-506},
          doi = {10.1111/j.1365-2966.2008.13805.x},
archivePrefix = {arXiv},
       eprint = {0808.1227},
 primaryClass = {astro-ph},
       adsurl = {https://ui.adsabs.harvard.edu/abs/2008MNRAS.391..481S}
}

@ARTICLE{tacchella2019,
       author = {{Tacchella}, Sandro and {Diemer}, Benedikt and {Hernquist}, Lars and {Genel}, Shy and {Marinacci}, Federico and {Nelson}, Dylan and {Pillepich}, Annalisa and {Rodriguez-Gomez}, Vicente and {Sales}, Laura V. and {Springel}, Volker and {Vogelsberger}, Mark},
        title = "{Morphology and star formation in IllustrisTNG: the build-up of spheroids and discs}",
      journal = {\mnras},
         year = 2019,
        month = aug,
       volume = {487},
       number = {4},
        pages = {5416-5440},
          doi = {10.1093/mnras/stz1657},
archivePrefix = {arXiv},
       eprint = {1904.12860},
 primaryClass = {astro-ph.GA},
       adsurl = {https://ui.adsabs.harvard.edu/abs/2019MNRAS.487.5416T}
}

@ARTICLE{tan2026,
       author = {{Tan}, Zhenlin and {Xie}, Lizhi and {Han}, Jiaxin and {Qiu}, Yisheng and {Fontanot}, Fabio and {De Lucia}, Gabriella and {Guo}, Qi and {Li}, Qingyang and {Zhou}, Jiale and {Jiang}, Wenkang and {Wang}, Xin and {He}, Feihong and {Jin}, Chichuan and {Jing}, Yipeng and {Li}, Ming and {Li}, Xiaodong and {Pei}, Wenxiang and {Wang}, Wenting and {Yang}, Xiaohu and {Yu}, Yu},
        title = "{A semi-analytical mock galaxy catalog for the CSST extragalactic surveys from the Jiutian simulations}",
      journal = {Science China Physics, Mechanics, and Astronomy},
         year = 2026,
        month = jan,
       volume = {69},
       number = {3},
          eid = {239512},
        pages = {239512},
          doi = {10.1007/s11433-025-2847-0},
archivePrefix = {arXiv},
       eprint = {2511.03281},
 primaryClass = {astro-ph.GA},
       adsurl = {https://ui.adsabs.harvard.edu/abs/2026SCPMA..6939512T}
}

@ARTICLE{taylor2011,
       author = {{Taylor}, Edward N. and {Hopkins}, Andrew M. and {Baldry}, Ivan K. and {Brown}, Michael J.~I. and {Driver}, Simon P. and {Kelvin}, Lee S. and {Hill}, David T. and {Robotham}, Aaron S.~G. and {Bland-Hawthorn}, Joss and {Jones}, D.~H. and {Sharp}, R.~G. and {Thomas}, Daniel and {Liske}, Jochen and {Loveday}, Jon and {Norberg}, Peder and {Peacock}, J.~A. and {Bamford}, Steven P. and {Brough}, Sarah and {Colless}, Matthew and {Cameron}, Ewan and {Conselice}, Christopher J. and {Croom}, Scott M. and {Frenk}, C.~S. and {Gunawardhana}, Madusha and {Kuijken}, Konrad and {Nichol}, R.~C. and {Parkinson}, H.~R. and {Phillipps}, S. and {Pimbblet}, K.~A. and {Popescu}, C.~C. and {Prescott}, Matthew and {Sutherland}, W.~J. and {Tuffs}, R.~J. and {van Kampen}, Eelco and {Wijesinghe}, D.},
        title = "{Galaxy And Mass Assembly (GAMA): stellar mass estimates}",
      journal = {\mnras},
         year = 2011,
        month = dec,
       volume = {418},
       number = {3},
        pages = {1587-1620},
          doi = {10.1111/j.1365-2966.2011.19536.x},
archivePrefix = {arXiv},
       eprint = {1108.0635},
 primaryClass = {astro-ph.CO},
       adsurl = {https://ui.adsabs.harvard.edu/abs/2011MNRAS.418.1587T}
}

@ARTICLE{toomre1972,
   author = {{Toomre}, A. and {Toomre}, J.},
    title = "{Galactic Bridges and Tails}",
  journal = {\apj},
     year = 1972,
    month = dec,
   volume = 178,
    pages = {623-666},
      doi = {10.1086/151823},
   adsurl = {http://adsabs.harvard.edu/abs/1972ApJ...178..623T}
}

@ARTICLE{vanderwel2014,
       author = {{van der Wel}, A. and {Franx}, M. and {van Dokkum}, P.~G. and {Skelton}, R.~E. and {Momcheva}, I.~G. and {Whitaker}, K.~E. and {Brammer}, G.~B. and {Bell}, E.~F. and {Rix}, H.-W. and {Wuyts}, S. and {Ferguson}, H.~C. and {Holden}, B.~P. and {Barro}, G. and {Koekemoer}, A.~M. and {Chang}, Yu-Yen and {McGrath}, E.~J. and {H{\"a}ussler}, B. and {Dekel}, A. and {Behroozi}, P. and {Fumagalli}, M. and {Leja}, J. and {Lundgren}, B.~F. and {Maseda}, M.~V. and {Nelson}, E.~J. and {Wake}, D.~A. and {Patel}, S.~G. and {Labb{\'e}}, I. and {Faber}, S.~M. and {Grogin}, N.~A. and {Kocevski}, D.~D.},
        title = "{3D-HST+CANDELS: The Evolution of the Galaxy Size-Mass Distribution since z = 3}",
      journal = {\apj},
         year = 2014,
        month = jun,
       volume = {788},
       number = {1},
          eid = {28},
        pages = {28},
          doi = {10.1088/0004-637X/788/1/28},
archivePrefix = {arXiv},
       eprint = {1404.2844},
 primaryClass = {astro-ph.GA},
       adsurl = {https://ui.adsabs.harvard.edu/abs/2014ApJ...788...28V}
}

@ARTICLE{vasiliev2022,
       author = {{Vasiliev}, Eugene and {Belokurov}, Vasily and {Evans}, N. Wyn},
        title = "{Radialization of Satellite Orbits in Galaxy Mergers}",
      journal = {\apj},
         year = 2022,
        month = feb,
       volume = {926},
       number = {2},
          eid = {203},
        pages = {203},
          doi = {10.3847/1538-4357/ac4fbc},
archivePrefix = {arXiv},
       eprint = {2108.00010},
 primaryClass = {astro-ph.GA},
       adsurl = {https://scixplorer.org/abs/2022ApJ...926..203V}
}

@ARTICLE{villalobos2013,
       author = {{Villalobos}, {\'A}. and {De Lucia}, G. and {Weinmann}, S.~M. and {Borgani}, S. and {Murante}, G.},
        title = "{An improved prescription for merger time-scales from controlled simulations}",
      journal = {\mnras},
         year = 2013,
        month = jun,
       volume = {433},
       number = {1},
        pages = {L49-L53},
          doi = {10.1093/mnrasl/slt056},
archivePrefix = {arXiv},
       eprint = {1303.2117},
 primaryClass = {astro-ph.CO},
       adsurl = {https://ui.adsabs.harvard.edu/abs/2013MNRAS.433L..49V}
}

@ARTICLE{wang2019,
       author = {{Wang}, Lan and {Xu}, Dandan and {Gao}, Liang and {Guo}, Qi and {Qu}, Yan and {Pan}, Jun},
        title = "{Comparing galaxy morphology in hydrodynamical simulation and in semi-analytic model}",
      journal = {\mnras},
         year = 2019,
        month = may,
       volume = {485},
       number = {2},
        pages = {2083-2091},
          doi = {10.1093/mnras/stz529},
archivePrefix = {arXiv},
       eprint = {1812.03645},
 primaryClass = {astro-ph.GA},
       adsurl = {https://ui.adsabs.harvard.edu/abs/2019MNRAS.485.2083W}
}

@ARTICLE{wilman2013,
       author = {{Wilman}, David J. and {Fontanot}, Fabio and {De Lucia}, Gabriella and {Erwin}, Peter and {Monaco}, Pierluigi},
        title = "{The hierarchical origins of observed galaxy morphology}",
      journal = {\mnras},
         year = 2013,
        month = aug,
       volume = {433},
       number = {4},
        pages = {2986-3004},
          doi = {10.1093/mnras/stt941},
archivePrefix = {arXiv},
       eprint = {1305.7163},
 primaryClass = {astro-ph.CO},
       adsurl = {https://ui.adsabs.harvard.edu/abs/2013MNRAS.433.2986W}
}

@ARTICLE{Wu2025,
       author = {{Wu}, Xinyi and {Zhu}, Ling and {Chang}, Jiang and {Zeng}, Guangquan and {Lei}, Yu},
        title = "{The effects of the orbital configurations of mergers on reshaping galaxy structures}",
      journal = {\aap},
         year = 2025,
        month = jul,
       volume = {699},
          eid = {A374},
        pages = {A374},
          doi = {10.1051/0004-6361/202554638},
archivePrefix = {arXiv},
       eprint = {2507.02051},
 primaryClass = {astro-ph.GA},
       adsurl = {https://ui.adsabs.harvard.edu/abs/2025A&A...699A.374W}
}

@ARTICLE{wuyts2011,
       author = {{Wuyts}, Stijn and {F{\"o}rster Schreiber}, Natascha M. and {van der Wel}, Arjen and {Magnelli}, Benjamin and {Guo}, Yicheng and {Genzel}, Reinhard and {Lutz}, Dieter and {Aussel}, Herv{\'e} and {Barro}, Guillermo and {Berta}, Stefano and {Cava}, Antonio and {Graci{\'a}-Carpio}, Javier and {Hathi}, Nimish P. and {Huang}, Kuang-Han and {Kocevski}, Dale D. and {Koekemoer}, Anton M. and {Lee}, Kyoung-Soo and {Le Floc'h}, Emeric and {McGrath}, Elizabeth J. and {Nordon}, Raanan and {Popesso}, Paola and {Pozzi}, Francesca and {Riguccini}, Laurie and {Rodighiero}, Giulia and {Saintonge}, Amelie and {Tacconi}, Linda},
        title = "{Galaxy Structure and Mode of Star Formation in the SFR-Mass Plane from z \raisebox{-0.5ex}\textasciitilde 2.5 to z \raisebox{-0.5ex}\textasciitilde 0.1}",
      journal = {\apj},
         year = 2011,
        month = dec,
       volume = {742},
       number = {2},
          eid = {96},
        pages = {96},
          doi = {10.1088/0004-637X/742/2/96},
archivePrefix = {arXiv},
       eprint = {1107.0317},
 primaryClass = {astro-ph.CO},
       adsurl = {https://ui.adsabs.harvard.edu/abs/2011ApJ...742...96W}
}

@ARTICLE{xie2017,
   author = {{Xie}, L. and {De Lucia}, G. and {Hirschmann}, M. and {Fontanot}, F. and 
	{Zoldan}, A.},
    title = "{H2-based star formation laws in hierarchical models of galaxy formation}",
  journal = {\mnras},
     year = 2017,
    month = jul,
   volume = 469,
    pages = {968-993},
      doi = {10.1093/mnras/stx889},
   adsurl = {http://adsabs.harvard.edu/abs/2017MNRAS.469..968X}
}

@ARTICLE{xie2018,
       author = {{Xie}, Lizhi and {De Lucia}, Gabriella and {Wilman}, David J. and
         {Fossati}, Matteo and {Erwin}, Peter and {Guti{\'e}rrez}, Leonel and
         {Kulkarni}, Sandesh K.},
        title = "{On the influence of environment on star-forming galaxies}",
      journal = {\mnras},
         year = "2018",
        month = "Nov",
       volume = {480},
       number = {3},
        pages = {3812-3825},
          doi = {10.1093/mnras/sty2131},
archivePrefix = {arXiv},
       eprint = {1808.01628},
 primaryClass = {astro-ph.GA},
       adsurl = {https://ui.adsabs.harvard.edu/abs/2018MNRAS.480.3812X}
}

@ARTICLE{xie2020,
       author = {{Xie}, Lizhi and {De Lucia}, Gabriella and {Hirschmann}, Michaela and {Fontanot}, Fabio},
        title = "{The influence of environment on satellite galaxies in the GAEA semi-analytic model}",
      journal = {\mnras},
         year = 2020,
        month = nov,
       volume = {498},
       number = {3},
        pages = {4327-4344},
          doi = {10.1093/mnras/staa2370},
archivePrefix = {arXiv},
       eprint = {2003.12757},
 primaryClass = {astro-ph.GA},
       adsurl = {https://ui.adsabs.harvard.edu/abs/2020MNRAS.498.4327X}
}

@ARTICLE{xie2025,
       author = {{Xie}, Lizhi and {De Lucia}, Gabriella and {Fossati}, Matteo and {Fontanot}, Fabio and {Hirschmann}, Michaela},
        title = "{The impact of ram pressure on cluster galaxies, insights from GAEA and TNG}",
      journal = {\aap},
         year = 2025,
        month = jun,
       volume = {698},
          eid = {A73},
        pages = {A73},
          doi = {10.1051/0004-6361/202553915},
archivePrefix = {arXiv},
       eprint = {2504.12863},
 primaryClass = {astro-ph.GA},
       adsurl = {https://ui.adsabs.harvard.edu/abs/2025A&A...698A..73X}
}





\appendix
\nolinenumbers

\section{Model calibration}
\label{appsec:calib}

\begin{figure}
    \centering
    \includegraphics[width=.8\linewidth]{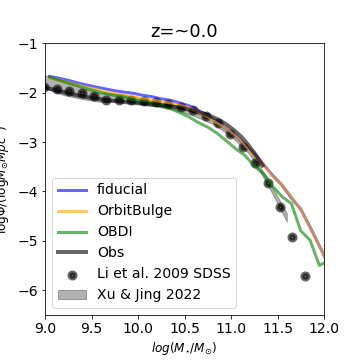}
    \includegraphics[width=.8\linewidth]{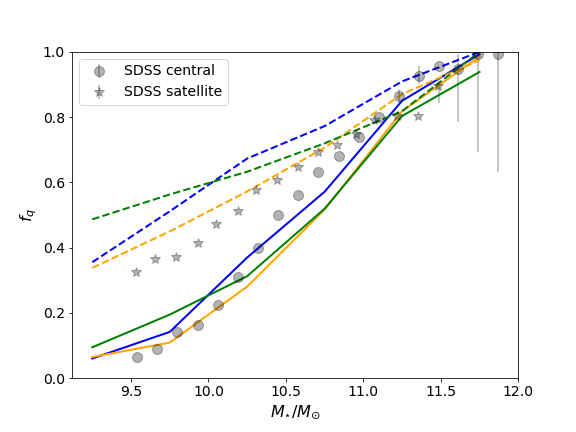}
    \caption{The top panles shows stellar mass functions at $z= 0$. Observational measurement of stellar mass function are from \citet{schreiber2020},\citet{li2009}, \citet{xu2022}. 
    The bottom panel shows quenched fractions at $z=0$. Central and satellite galaxies are shown by solid and dashed lines. Observed quenched fractions are estimates based on SDSS data published in \citet{hirschmann2014}. }
    \label{fig:calibration}
\end{figure}

We calibrate the new models to reproduce the stellar mass functions  and quenched fraction for central and satellite galaxies at $z=0$. The results are shown in Figure~\ref{fig:calibration}. The parameters are listed in Table~\ref{tab:modelParam}. 

\begin{table*}[]
    \small
    \centering
    \begin{tabular}{lcccc}
    \hline
           Parameter & meaning & fiducial & \ob & \obdi \\
         \hline
          $\epsilon_{\rm reheat}$ & Reheating efficiency &  0.28 & 0.28 & 0.18\\
          $\epsilon_{\rm eject}$  & Ejection efficiency &   0.1 & 0.15 & 0.2\\
          $\gamma_{\rm reinc}$ & Reincorporation efficiency & 0.99 & 0.99 & 0.63\\
          $\kappa_{\rm radio}/10^{-5}$ & Hot gas Black Hole accretion efficiency & 1.36 & 2 & 1.36\\
          $f_{\rm lowJ}/10^{-3}$ & Cold gas angular momentum loss efficiency & 3.18 & 3.18 & 2.5\\
          $f_{\rm BH}/10^{-3}$  & Black hole accretion rate from reservoir & 0.11 & 0.3 & 0.15\\
          $\epsilon_{\rm qw}/100$ & Quasar wind efficiency & 4.86 & 4.86 & 1\\
          $f_{\rm cen}/10^{-3}$  & Fraction of ISM added to the black hole reservoir  & 3.39 & 3.39 & 2.0\\ 
          & as a consequence of AGN-driven outflow & & &\\
          rpstime & timescale of ram pressure stripping & 444 & 444 & 650 \\
          $C_{\rm rad}$ & fraction of energy radiated via mergers & 0 & 1 & 2 \\
    \hline    
    \end{tabular}
    \caption{Parameter calibration}
    \label{tab:modelParam}
\end{table*}

\section{Bulges in the MSII}
\label{appsec:MSII}

\begin{figure*}
    \centering
    \includegraphics[width=1.\textwidth]{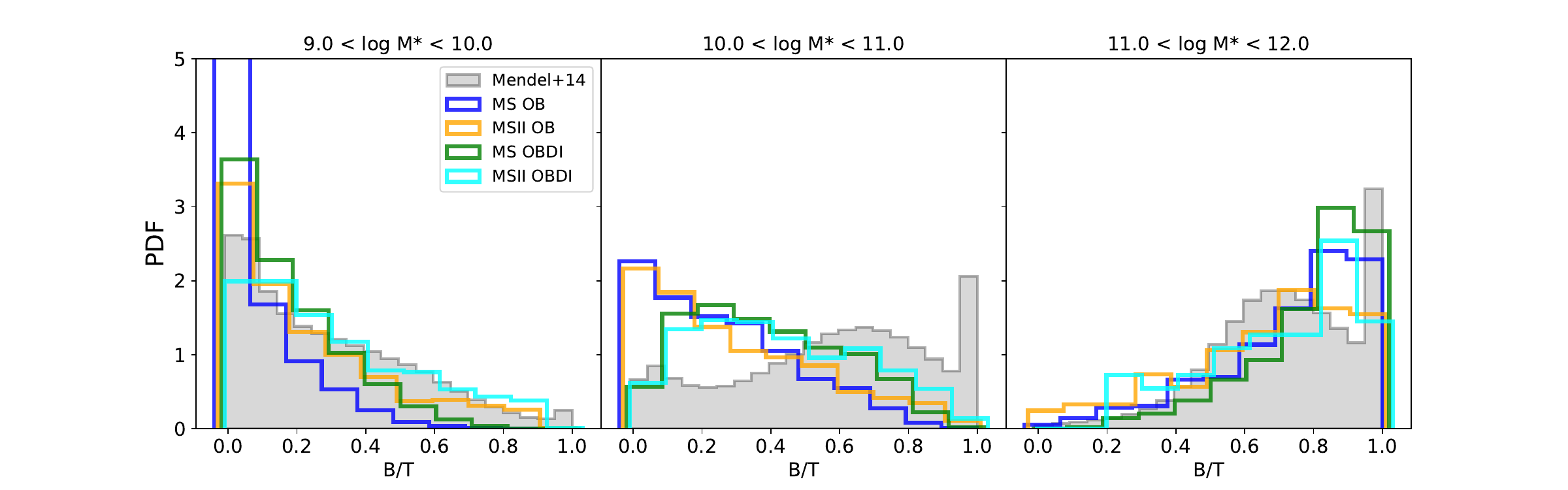}
    \caption{The distribution of $\btt$ ratio for galaxies from the MSII.}
    \label{fig:btt_MSII}
\end{figure*}

Motivated to check the effect of resolution, we apply the modified model to the MSII, and compare the $\btt$ distribution of galaxies with the results of MS. The result is shown in Fig~\ref{fig:btt_MSII}. The low-mass galaxies in MSII show higher $\btt$ fractions compared to those in the MS, and are in better agreement with SDSS. Massive galaxies are not affected by resolution. 

\section{Impact of Gas disk instability}
\label{appsec: DI}

\begin{figure*}
    \centering
    \includegraphics[width=1.\textwidth]{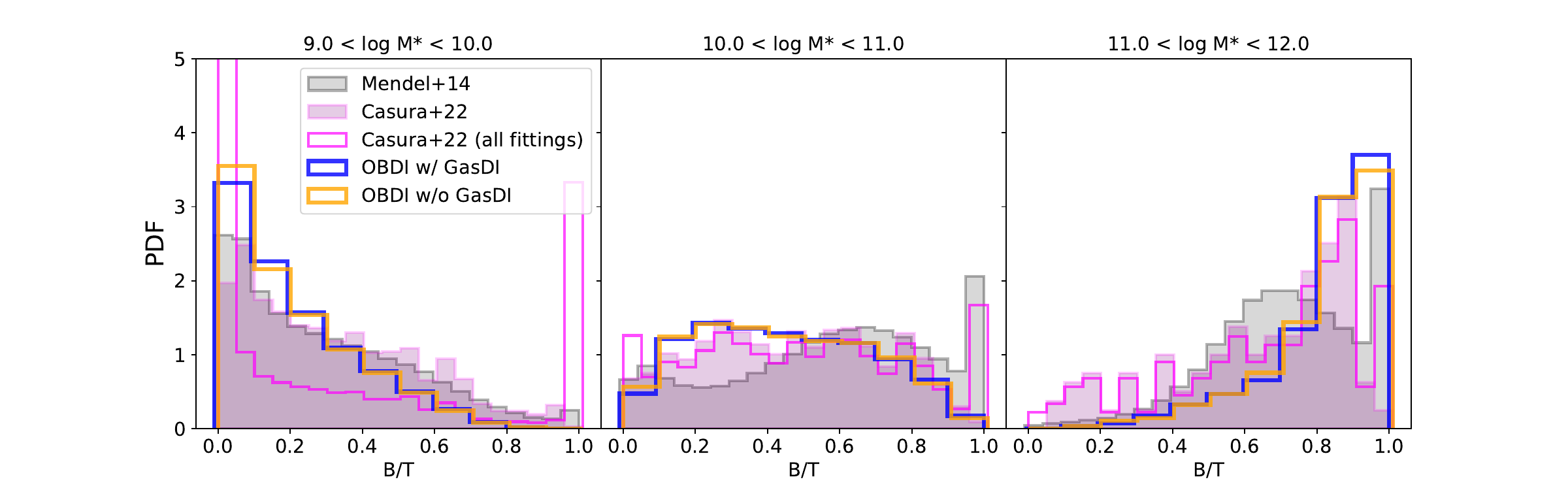}
    \caption{The distribution of $\btt$ ratio predicted by \obdi runs with/without the gas instability prescription.}
    \label{fig:btt_gasdi}
\end{figure*}

In Fig~\ref{fig:btt_gasdi} we quantify the relative contribution of the instability in gaseous and stellar disc to bulge formation by switching off gaseous instability. The gaseous instability slightly increases $\btt$ ratio for galaxies in all stellar mass bins, but the impact is minor compared to the contribution of stellar disc instability.

In the main text we only select galaxies from \citet{casura2022} that prefer two-component fitting, and have excluded more than half of the sample. To show how our selection could bias the overall distribution of $\btt$, The magenta histograms in Fig~\ref{fig:btt_gasdi} show the distribution of all galaxies with reliable fit. For galaxies with single  S\'ersic fit, we assign galaxies with $n <2$ as disc galaxies with $\btt = 0$, and those with $n> 2$ as elliptical galaxies with $\btt = 1$. 

\label{lastpage}
\end{document}